\documentclass[prb,aps,twocolumn,superscriptaddress]{revtex4-2}

\usepackage{amsmath,amssymb,amsfonts,bbm,color,graphicx}
\usepackage[caption=false]{subfig}

\usepackage[utf8]{inputenc}
\usepackage[T1]{fontenc}

\usepackage[unicode=true,colorlinks=true]{hyperref}
\hypersetup{linkcolor=blue,citecolor=blue,urlcolor=blue}

\usepackage{dcolumn}
\usepackage{bm}
\usepackage{dsfont}
\usepackage{soul}
\usepackage[usenames,dvipsnames]{xcolor}
\usepackage{wasysym}
\usepackage{enumitem}

\usepackage{braket}

\newcommand{\iu}{\mathrm{i}} 
\newcommand{\eu}{\mathrm{e}} 
\newcommand{\hc}{\mathrm{h.c.}} 

\newcommand{\bvec}[1]{\bm{#1}}

\newcommand{\Ztwo}{\mathbb{Z}_2}
\newcommand{\Uone}{\mathrm{U(1)}}
\newcommand{\SUtwo}{\mathrm{SU(2)}}

\begin{document}

\title{Emergent magnetism and spin liquids in an extended Hubbard description of moir\'e bilayers\\}

\author{Zhenhao Song}
\affiliation{Department of Physics, University of California, Santa Barbara, California 93106, USA}
\author{Urban~F.~P.~Seifert}
\affiliation{Institute for Theoretical Physics, University of Cologne, 50937 Cologne, Germany}
\affiliation{Kavli Institute for Theoretical Physics, University of California, Santa Barbara, California 93106-4030, USA}
\author{Leon Balents}
\affiliation{Kavli Institute for Theoretical Physics, University of California, Santa Barbara, California 93106-4030, USA}
\affiliation{French American Center for Theoretical Science, CNRS, KITP, Santa Barbara, California 93106-4030, USA}
\affiliation{Canadian Institute for Advanced Research, Toronto,  Ontario, Canada}
\author{Hong-Chen Jiang}
\affiliation{Stanford Institute for Materials and Energy Sciences, SLAC National Accelerator Laboratory, Menlo Park, California 94025, USA}

\begin{abstract}
Motivated by twisted transition metal dichalcogenides (TMDs), we study an extended Hubbard model with both on-site and off-site repulsive interactions, in which  Mott insulating states with concomitant charge order occur at fractional fillings. To resolve the charge ordering as well as the fate of the local moments formed thereby, we perform large-scale density matrix renormalization group calculations on cylindrical geometries for several filling fractions and ranges of interaction strength. Depending on the precise parameter regime, both antiferromagnetically ordered as well as quantum-disordered states are found, with a particularly prominent example being a quantum spin liquid-type ground state on top of charge-ordering with effective Kagom\'e geometry.  We discuss the different mechanisms at play in stabilizing various electronic and magnetic states.  The results suggest that moiré TMDs are a promising venue for emergent quantum magnetism of strongly interacting electrons.
\end{abstract}


\maketitle


\section{Introduction}
The Hubbard model provides perhaps the minimal example of a strong correlation Hamiltonian: simple to formulate, but difficult to solve.
It is believed to account for various exotic phenomena such as quantum magnetism, high-$T_c$ superconductivity, and quantum spin liquids \cite{damascelli03,lee06,lee05,savary17,arovas21,jiang20}. 
The effort expended by theorists to understand these phenomena is enormous and difficult to measure.

Hubbard models by definition involve only on-site Coulomb interaction,
making the effects of correlation progressively weaker as the density
of electrons is moved from half-filling. More generally, off-site
interactions can induce correlation physics ``centered'' away from
half-filling, for example the formation of charge ordered states (also
called Wigner crystals).  For such phenomena, extended Hubbard models,
which include some Coulomb repulsion between further neighbors, offer
a similarly minimalist formulation for theoretical investigation.
Extended Hubbard models have been studied, albeit to a much lesser
extent, in the theory literature \cite{zhang1989extended,PhysRevB.95.115149,aichhorn2004charge,sandvik2004ground,PhysRevLett.113.246405}.

A particular recent motivation to study extended Hubbard models arises from their realization in transition metal dichalcogenides (TMD) bilayers exhibiting a nanometer scale moir\'e pattern.  A microscopic theory beginning with a continuum model\cite{wu18,pan20a} can, when the topmost moir\'e band has trivial topology, be reduced to an extended Hubbard model living on the triangular moir\'e lattice, both in hetero-bilayer\cite{wu18} or twisted homobilayer\cite{pan20a} TMD structures.  Both Mott insulating phases at half-filling \cite{wang20,limak21} and Wigner crystals were also observed at other filling fractions\cite{regan20,liwang21}. The latter are only explained by the presence of substantial off-site interactions.  

Theoretically, the appropriate extended Hubbard model on the
triangular lattice has been studied by several methods.   Hartree-Fock
calculations predicted several competing magnetically ordered
states \cite{pan20b,hu21}. Mean field calculations using parton
decompositions were carried out at various
fillings \cite{sszb23,rossi23}, which suggested several spin liquid
states as alternatives to magnetic order. Both of these types of
studies are mean-field approximations, and in particular the parton
ones are quantitatively unreliable and should be regarded as a guide
for more accurate approaches.   Several more precise studies do exist,
but with more limited scope: Motruk \textit{et al.} derived an
effective Heisenberg model in the strong-coupling limit of an
effective Hubbard model at 3/4 filling, and mapped out phase diagrams
of the effective spin model \cite{motruk23}. Zhou \textit{et al.} used
large-scale density matrix renormalization group (DMRG) simulations to study the transition out of a Fermi
liquid into a generalized Wigner crystal at 1/3 filling, finding
120$^\circ$ magnetic ordering \cite{zhou2024quantum}, and a recent
study reports canted antiferromagnetism on top of a honyecomb
generalized Wigner crystal at 2/3 filling \cite{biborski24}.

Here, we aim to develop a more extensive picture of possible phases that may emerge at different fillings and spatial extent of repulsive interactions by means of large-scale 
\emph{direct} DMRG simulation of the extended Hubbard model.
Our simulations are motivated by predictions from our previous slave-rotor mean-field theory calculations \cite{sszb23}: we focus in particular on commensurate 4/3, 5/3, 5/4 filling factors, where we expect stable charge crystallization to occur due to relatively short-ranged (up to second-nearest neighbor) Coulomb repulsions.
We complement our numerical results by controlled analytical arguments.

The paper is organized as follows. 
In Sec.~\ref{Sec:model}, we briefly introduce the model Hamiltonian, as well as the setup of our DMRG calculations. 
Sec.~\ref{Sec:Phase diagram} provides a schematic phase diagram and summary of our results.  
Sec.~\ref{Sec:order} is devoted to magnetically ordered states and which interactions stabilize these.
In Sec.~\ref{Sec:disorder}, we examine two magnetically disordered states. We characterize these states in details and discuss underlying mechanisms.
We summarize our results in Sec.~\ref{Sec:Summary}.

\section{Model and methods} \label{Sec:model}

\subsection{Hubbard model for moiré transition metal dichalcogenides}

We consider an extended Hubbard model on the triangular lattice, motivated by effective models for moiré TMD heterobilayer and twisted homobilayer heterostructures \cite{wu18,pan20a}. Note that we are agnostic towards particular material realizations.
Specifically, the Hamiltonian of interest consists of a hopping term $\mathcal{H}_t$ and extended repulsive interactions $\mathcal{H}_\mathrm{int}$, 
\begin{subequations}\begin{align}
    \mathcal{H} &= \mathcal{H}_t + \mathcal{H}_{\text{int}} \label{eq:h-model} \\
    \mathcal{H}_t &= -t \sum_{\sigma}\sum_{\langle i j \rangle} \left(c_{i,\sigma}^\dagger c_{j,\sigma}^{\vphantom{\dagger}} + \hc \right) \label{eq:h-t} \\
    \mathcal{H}_{\text{int}} &= \sum_{ij}U_{ij}(n_i-1)(n_j-1) \label{eq:h-int}
\end{align}\end{subequations}
where $\sigma = \uparrow,\downarrow$ indexes the pseudospin-1/2 degree of freedom, and $n_i = n_{i,\uparrow} + n_{i,\downarrow}$ denotes the total number of particles per site.
For simplicity, we truncate repulsive interactions beyond second-nearest neighbor, and we define $U_{ij} = U/2,V/2,V'/2$,
for onsite ($i=j$), nearest neighbor and second-nearest neighbor interactions, respectively (the 1/2 factor of $V$ and $V'$ compensates the double counting of pairs). Note that in moiré TMD heterostructures, the spatial range of repulsive interactions can be controlled by an adjacent screening gate \cite{wu18,mak22}.

In Eq.~\eqref{eq:h-int}, we have rewritten the interaction terms in a manifestly particle-hole symmetric manner, which is related to the commonly used onsite repulsion term $n_{i,\uparrow}n_{i,\downarrow} = (n_i -1)^2/2 + n_i/2 -1/2$ by a redefinition of the chemical potential and a constant energy shift. 
We omit the constant as well as the chemical potential term, since we perform calculations at fixed fillings.
In the following, we will focus on filling factors $\bar{n} = N^{-1} \sum_i n_i \geq 1$ for simplicity. While the Hamiltonian \eqref{eq:h-model} does not possess particle-hole symmetry, we expect that most of our results are \emph{qualitatively} applicable also for $\bar{n} \leq 1$.

\subsection{Density-matrix renormalization group}
\label{sec:dmrg}
\begin{figure}[ht]
	\centering
	\includegraphics[width=\linewidth]{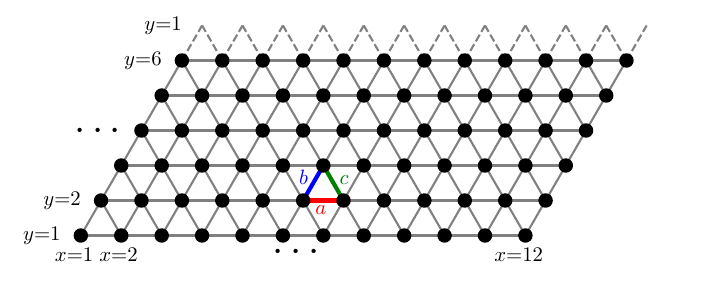}
	\caption{A schematic XC6 geometry with $L_x=12$ and $L_y=6$. $L_x=48$ is used in the actual DMRG simulation. The cylinder has a periodic boundary condition along the direction $\hat{y}=(1/2,\sqrt{3}/2)$, and an open boundary condition along $\hat{x}=(1,0)$ direction. The colored three bonds ($a$, $b$, $c$) denote three different orientations.
 \label{fig:XC6}
 }
\end{figure}

To find the ground state of $\mathcal{H} = \mathcal{H}_t + \mathcal{H}_\mathrm{int}$, we perform DRMG simulations on a XC6 geometry cylinder shown in Fig.~\ref{fig:XC6}, with one of the triangular lattice's bonds being parallel to the $\hat{x}$ direction \cite{Yan2011}.
Along the second primitive translation direction $\hat{y}=(1/2,\sqrt{3}/2)$ we use periodic boundary conditions with a width of $L_y=6$, to accommodate the anticipated honeycomb and Kagom\'e charge orders \cite{sszb23}.
Along the $\hat{x}$ direction, there is an open boundary condition with a length up to $L_x=48$.

We study the single band extended Hubbard model Eq.~\eqref{eq:h-model} at fixed fillings of $\bar{n}=$4/3, 5/3 and 5/4. The parameters are set as $t=1$, $U=12t$, $V=U/4$ \cite{pan20a,xu20,tingxin21}, and we use $V^{\prime}=0$ and $V/\sqrt{3}$ to model short-ranged and longer-ranged interactions, respectively.
We perform both $\Uone$ and $\SUtwo$ DMRG simulations, which impose $\Uone$ charge conservation and $\Uone$ ($\SUtwo$, respectively) spin rotation symmetries explicitly, and compare the energies of the two.
The $\Uone$ DMRG simulation allows for the emergence of ground states with broken spin-rotation symmetry (finite out-of-plane spin components $\langle S^z_i \rangle \neq 0$) as the DMRG code enforces only a $\Uone$ symmetry of spin rotations along that $\hat{z}$ axis.
In contrast, the $\SUtwo$ DMRG simulation preserves the full $\SUtwo$ spin rotation symmetry of $\mathcal{H}$ in Eq. \eqref{eq:h-model} and therefore only converges to $\SUtwo$-symmetric states.

In the $\Uone$ DMRG simulation, we keep up to $m=40\,000$ number of states with a typical truncation error $\epsilon\approx 3\times 10^{-6}$. In the $\SUtwo$ DMRG simulation, we keep the bond dimension of $\SUtwo$ multiplets up to $D=30\,000$ with a typical truncation error $\epsilon\approx 10^{-6}$, which are equivalent to $m\approx 88\,000$ U(1) states.
The convergence has been checked under different initial wavefunctions, different pinning fields, and different number of states.

\section{Overall Phase Diagram} \label{Sec:Phase diagram}
\begin{figure}[ht]
	\centering
	\includegraphics[width=\linewidth]{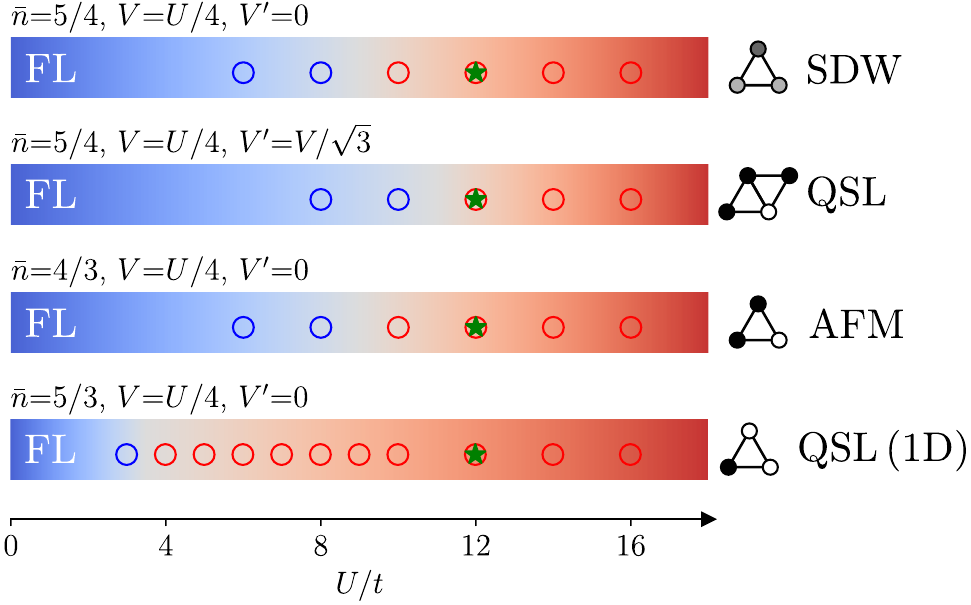}
	\caption{A schematic phase diagram in terms of interaction strengths $U/t$ and filling $\bar{n}$. The circles indicate where we have done DMRG simulations. Red circles indicate an incompressible states with concomitant charge ordering, while blue symbols denote metallic state that perserves all lattice symmetries. The underlying shading illustrates \emph{speculated} metallic and insulating regions in the phase space. Green stars are data points where we perform carefully converged DMRG calculations to investigate the precise nature of respective ground states, and are discussed in detail in the following sections.} 
 \label{fig:phase diagram} 
\end{figure}

We first sketch the phase diagram in Fig.~\ref{fig:phase diagram}, in terms of interaction strength $U/t$ for different fillings and range of repulsive interactions. Circle symbols denote data points where we have perfomed DMRG calculations. We encode metallic states with blue symbols, and insulating regions are shaded in red, accompanied by a modulation in the charge density (see below).
In our DMRG simulation, distinct phases are characterized by evaluating a range of physical quantities, such as charge and spin density profiles, and correlation functions including spin, charge, superconducting pair-field, dimer and spin chirality correlations.

We succinctly highlight salient features of the overall phase diagram:

\begin{enumerate}
    \item For all parameter sets considered, at sufficiently large $U/t$, the model forms an incompressible state with a concomitant translation-symmetry breaking charge density.  There are noticeable variations in the critical interaction strength. In particular, we observe that for the rather dense filling factor $\bar{n}=5/3$ relatively weak interactions are sufficient for driving the system into an interaction-induced insulating state. We further note that at filling factor $\bar{n}=5/4$, switching on longer-ranged repulsive interactions leads to an \emph{increase} in critical interaction strength required to induce insulating behaviour.
    \item All interaction-driven insulating states exhibit a spatially modulated charge density, thereby spontaneously breaking the translational/rotational symmetries of the underlying triangular lattice. Here, we distinguish between the formation of generalized Wigner crystals which are characterized by an (approximate) integer number of charges per site $\langle n_i \rangle \approx 0,1,2$ and the formation of a spatially weakly modulated charge-density $\langle n_i \rangle = \bar{n} + \rho(\bvec{r}_i)$ with $|\rho| \ll \bar{n}$. We find generalized Wigner crystals for short-ranged interactions ($V'=0$) for $\bar{n}=4/3$ with an emergent honeycomb lattice of singly-occupied sites, and at $\bar{n}=5/3$ with an emergent triangular lattice of singly occupied sites. In contrast, we observe weak charge-density wave ordering (with lattice momentum $\bvec{K}$) at $\bar{n}=5/4$. At this filling, a generalized Wigner crystal with an emergent Kagom\'e lattice of singly occupied sites is stabilized upon switching on longer-ranged repulsive interactions ($V' = V/\sqrt{3}$).
    \item Depending on the nature of charge-ordering in the insulating state, both magnetically ordered as well as quantum-paramagnetic states are found to emerge. Our numerical simulations show that the generalized Wigner crystal at $\bar{n}=4/3$ with an effective honeycomb geometry of singly occupied sites, exhibits zigzag antiferromagnetic order of local moments, while the triangular lattice Wigner crystal at $\bar{n}=5/3$ does not exhibit bulk magnetic order. The CDW-ordered state at $\bar{n}=5/4$, $V'=0$ is found to posses a spin-density wave order, while the Kagom\'e Wigner crystal stabilized by longer-ranged interactions (at $\bar{n}=5/4$, $V'=V/\sqrt{3}$) features vanishing magnetic moments in the ground state. Taken together with the observed absence of chiral and dimer ordering, this may serve as an indication of a quantum spin liquid ground state.
\end{enumerate}

In the following, we discuss above observations pertaining the four parameter sets we studied in more detail. We organize these in terms of a finite/vanishing magnetic order, and discuss underlying magnetic interactions and mechanisms for the stabilization of these states in detail: Sec.~\ref{Sec:order} is concerned with the zigzag local-moment antiferromagnetism at $\bar{n}=4/3$ and the spin-density wave order found for $\bar{n}=5/4$, $V'=0$, while we refer the reader to Sec.~\ref{Sec:disorder} for an in-depth discussion of the quantum-disordered states at $\bar{n}=5/3$ and $\bar{n}=5/4$ ($V'=V/\sqrt{3}$).

\section{Magnetically ordered states} \label{Sec:order}

\subsection{Local moment zigzag antiferromagnet at $\bar{n}=4/3, V'=0$} \label{sec:R43}
\begin{figure}[t]
	\centering
	\includegraphics[width=0.9\linewidth]{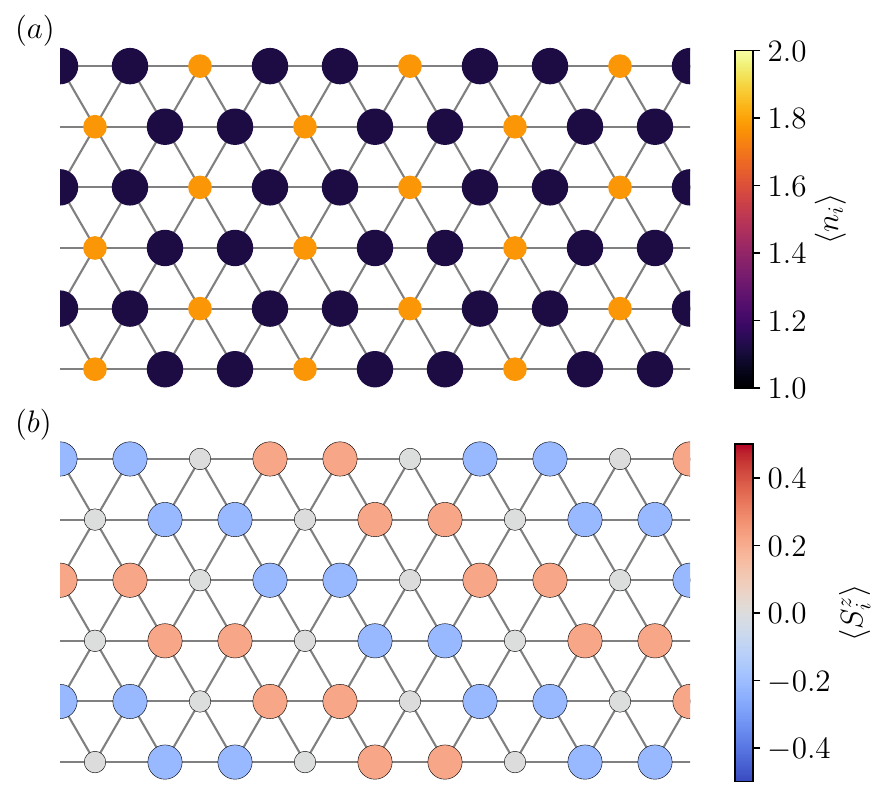}
	\caption{Magnetically ordered state at filling $\bar{n}=4/3$ with $V'=0$. (a) honeycomb charge order. (b) zigzag magnetic order. 
    Only the central region of the $N=48 \times 6$ lattice is shown to emphasize the bulk property. 
    Here, $U=12t$, $V=U/4$ and $V'=0$ with $\Uone$ DMRG simulation of bond dimension $m=25,000$.
 \label{fig:R43} 
 }
\end{figure}

We first consider the case of $\bar{n}=4/3$, where a classical analysis (setting the hopping $t = 0$, whereupon $\mathcal{H}$ in Eq.~\eqref{eq:h-model} corresponds to a classical lattice gas) as well as our slave-rotor mean-field theory \cite{sszb23} predicts that charges form an effective honeycomb lattice of singly-occupied sites, with the centers of the hexagons hosting doubly occupied sites.

\begin{figure}[t]
	\centering
	\includegraphics[width=0.8\linewidth]{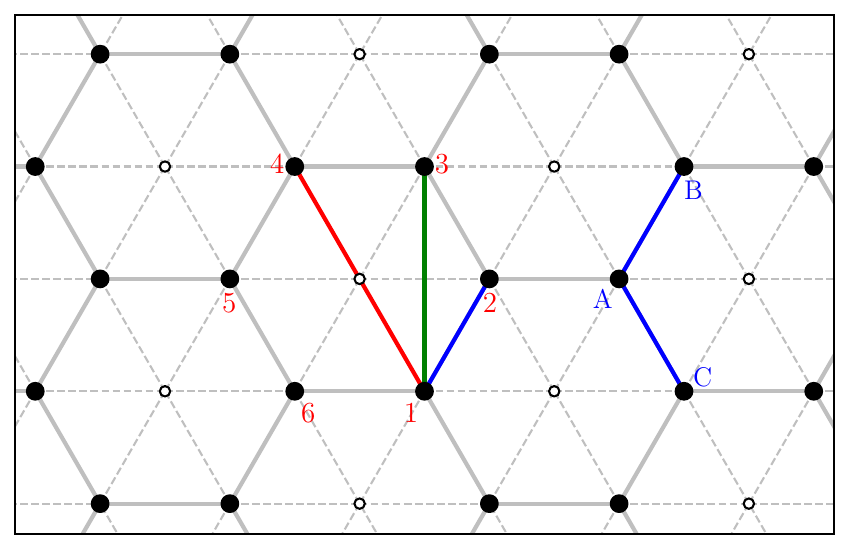}
	\caption{Effective honeycomb sublattice of the Wigner crystal at filling $\bar{n}=4/3$ for $V^{\prime}=0$. Filled circles represent singly occupied sites while empty ones represent doubly occupied sites. The blue, green and red lines denote nearest, next nearest, and third nearest neighbor pairs, respectively. Bond $\braket{\text{12}}$ and $\braket{\text{AB}}$ or $\braket{\text{AC}}$ are nearest but non-overlapping bonds, which can be connected by one single (honeycomb) bond.
 \label{fig:honeyc sublat}}
\end{figure}

Our DMRG results for this case are shown in Fig.~\ref{fig:R43}, where we exhibit only the middle region of the $N=48\times6$ cylinder in order to illustrate bulk physics. We do so in the remainder of this paper, as well. 
The observed bulk honeycomb charge crystal [Fig.~\ref{fig:R43}(a)] is in accord with our classical/mean-field calculations.
In Fig.~\ref{fig:R43}(b), the out-of-plane spin component $\langle S^z_i \rangle$ as obtained from our $\Uone$-DMRG simulation is shown: we find that singly occupied sites exhibit collinear zigzag antiferromagnetic order.
As the charge crystallization leads to the emergence of local moments with unit occupancy, in the limit of $U\gg t$ we may fix a static (``classical'') charge configuration and perform standard $t/U$ perturbation theory to derive an effective Heisenberg model for spin-spin interactions of local moments on top of the background charge crystals.
Crucially, we also account for processes which involve doubly occupied sites (at the centers of the hexagons of the honeycomb lattice). These processes do not occur if one were to consider just a honeycomb lattice Hubbard model.
The effective Hamiltonian up to fourth order in $t/U$ is found to be of the form
\begin{align}
    \mathcal H_{\text{eff}}
    &=J_1\sum_{\langle ij\rangle}\vec{S}_i\cdot \vec{S}_j
    +J_2\sum_{\langle \langle ij \rangle \rangle}\vec{S}_i\cdot \vec{S}_j
    \notag \\
    &+J_3\sum_{\langle \langle \langle ij \rangle \rangle \rangle}\vec{S}_i\cdot \vec{S}_j
    +K\sum{}'_{\langle ij\rangle,\langle kl\rangle}(\vec{S}_i\cdot \vec{S}_j)(\vec{S}_k\cdot \vec{S}_l). \label{eq:eff_H}
\end{align}
Here, $\langle ij \rangle$, $\langle \langle ij \rangle \rangle$, $\langle \langle \langle ij \rangle \rangle \rangle$ denote nearest, second-nearest and third-nearest neighbors on the effective honeycomb sublattice, respectively.
The fourth term in Eq.~\eqref{eq:eff_H} corresponds to a four-spin interaction between nearest, but non-overlapping honeycomb bonds, see Fig.~\ref{fig:honeyc sublat} for an illustration.
We give explicit expressions for the coefficients in terms of $t, U$ and $V$ in Appendix.~\ref{app:perturb}.
For the particular values of $U=12t$ and $V=U/4$, we obtain 
\begin{align}
    &J_1\approx -0.092 t, \quad J_2 \approx 0.016 t \quad J_3 \approx 0.022 t
    \notag \\
    &\text{and} \quad K \approx -0.00125 t.
    \label{eq:num_J}
\end{align}
We conclude that the nearest-neighbor exchange $J_1 < 0$ is in fact ferromagnetic, and the second-nearest neighbor and third-nearest neighbor exchanges are antiferromagnetic.
The four-spin interaction $K$ is one order-of-magnitude smaller than $J$ and will therefore be ignored in the following.
As a result, the effective spin Hamiltonian for the local moments  becomes a $J_1$-$J_2$-$J_3$ Heisenberg model on a honeycomb lattice.
Since we observe a (semiclassical) ordered magnetic state in our simulations, we may turn to a classical analysis of this effective spin model. Previous works find that the ground state of the classical $J_1$-$J_2$-$J_3$-Heisenberg model on the honeycomb lattice is indeed collinear zigzag antiferromagnetically ordered, and stable to quantum corrections \cite{rastelli79,fouet01}.
These results are in full agreement with our observations from numerical DMRG simulations, and we conclude that the antiferromagnetic order at $\bar{n} = 4/3$ is a prototypical example of local-moment antiferromagnetism stabilized by exchange interactions.

\subsection{Spin-density wave instability at $\bar{n}=5/4, V'=0$}
\label{sec:R54_Vp0}
\begin{figure}[ht]
	\centering
	\includegraphics[width=0.9\linewidth]{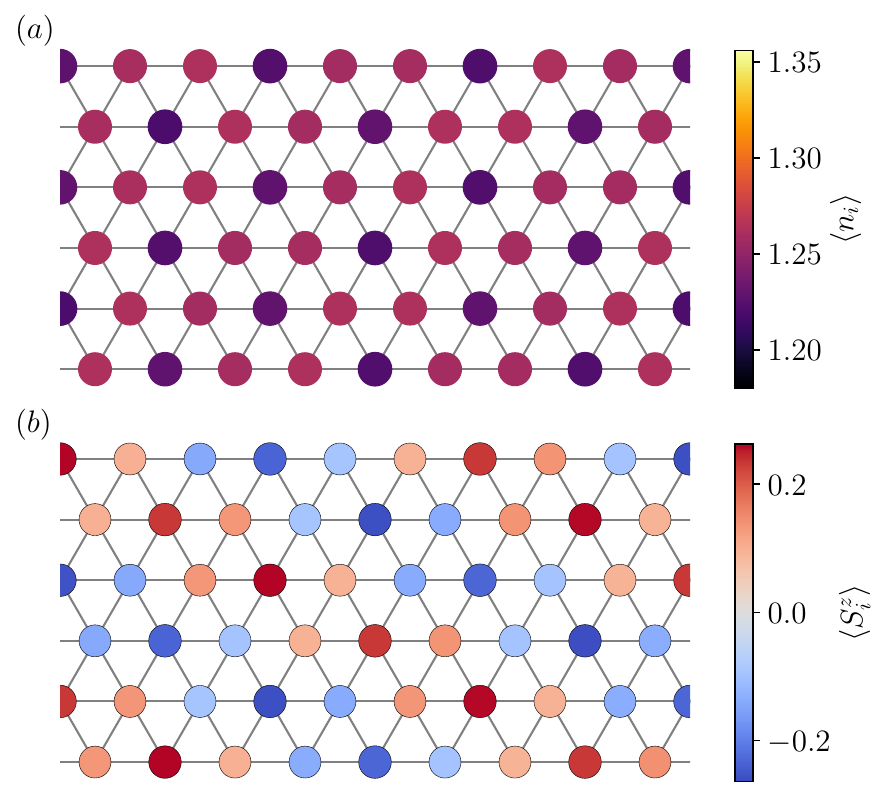}
	\caption{Magnetically ordered state at filling $\bar{n}=5/4$ with $V'=0$. (a) weak charge density wave order. (b) weak spin density wave order. 
    Only the central region of the $N=48 \times 6$ lattice is shown to emphasize the bulk property.
    Here, $U=12t$, $V=U/4$ and $V'=0$ with $\Uone$ DMRG simulation of bond dimension $m=40,000$.
 \label{fig:R54Vp0} 
 }
\end{figure}

We now turn to the filling factor $\bar{n} = 5/4$ with only onsite $U\neq 0$ and nearest neighbor $V\neq 0$ finite, and vanishing longer-ranged interactions $V'=0$.
In the classical limit ($t=0$),  $\mathcal{H}_\mathrm{int}$ does not admit a thermodynamically stable charge crystal state at this filling, as also supported by our previous slave-rotor mean-field theory results which point towards a paramagnetic metallic state for this parameter regime \cite{sszb23}.

\begin{figure}[t]
  \centering
  \includegraphics[width=.95\linewidth]{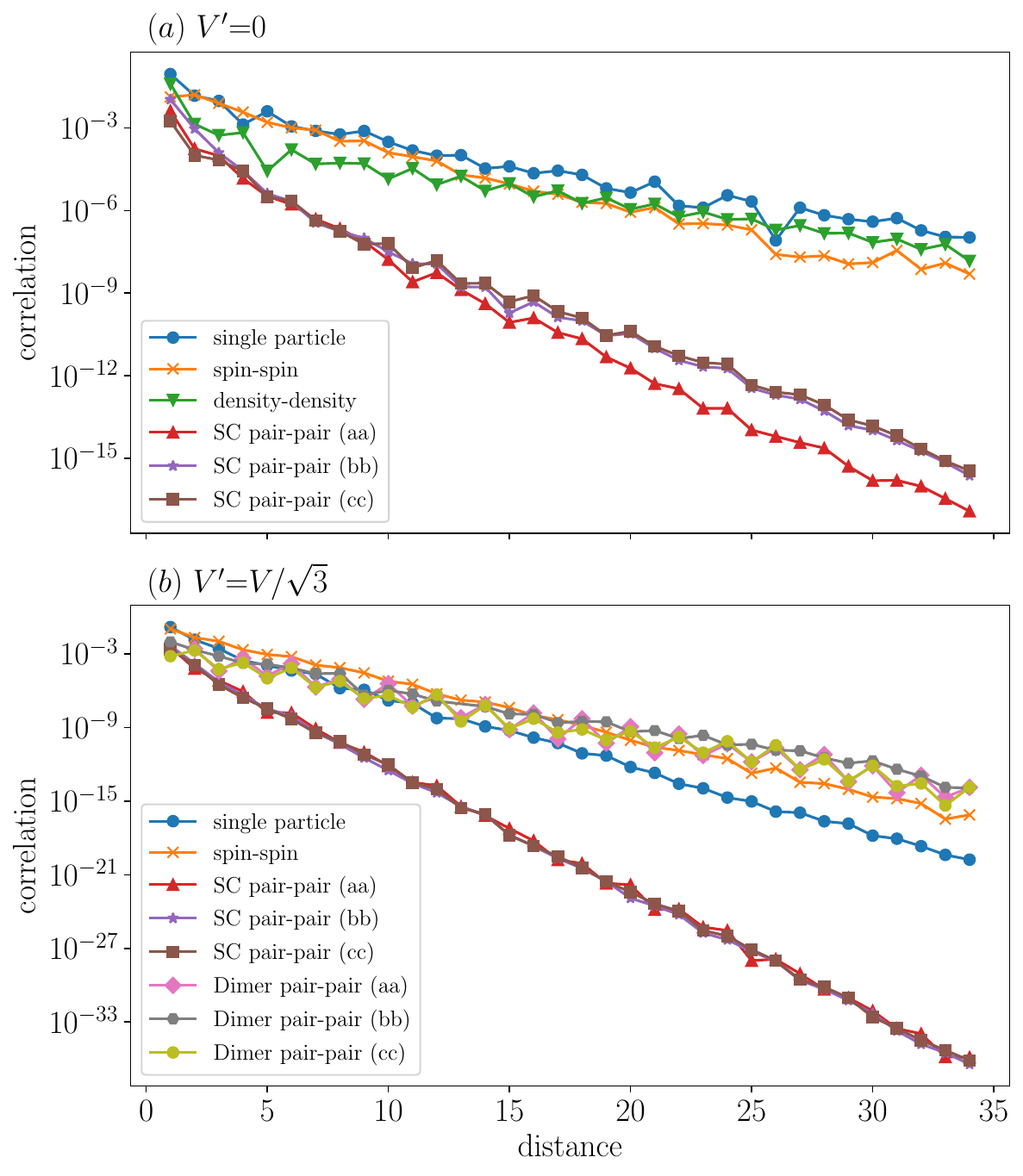}
  \caption{Various correlators at filling $\bar{n}=5/4$ with (a) $V'=0$ and (b) $V'=V/\sqrt{3}$ as a function of distance. 
  $a,b,c$ correspond to the three bond orientations, see Fig.~\ref{fig:XC6}.
  All correlations exhibit an exponential decay. 
  }
  \label{fig:combined corr}
\end{figure}

However, as displayed in the phase diagram in Fig.~\ref{fig:phase diagram}, our DMRG simulations indicate that beyond some critical interaction strength, the system becomes insulating as evidenced by an exponential decay of the single-particle and pair-pair correlations, shown in Fig.~\ref{fig:combined corr}a. 
In addition to the appearance of these indications of a charge gap, we observe symmetry-breaking  charge-density wave (CDW) and spin-density wave ordering (SDW).
We present bulk measurements of the site-resolved occupation number $\langle n_i \rangle$ and spin expectation value $\langle S^z_i \rangle$ in Fig.~\ref{fig:R54Vp0}.
Notably, we find that the amplitude $|\rho (\bvec{r}_i)|$ of the modulation of the charge density $\langle n_i \rangle = \bar{n} + \rho(\bvec{r}_i)$ around its mean $\bar{n}$ to be an order of magnitude smaller than the modulation of the spin density $\langle S^z_i \rangle = 0 + S^z(\bvec{r}_i)$.

The substantially larger size of the spin order suggests that it drives the charge one.  This is consistent with symmetry considerations.  We observe from Fig.~\ref{fig:R54Vp0} the wavevectors $\bvec{Q}_\mathrm{c}$ and $\bvec{Q}_\mathrm{s}$ of the charge-density and spin-density wave order parameters are
\begin{align}
    \bvec{Q}_\mathrm{c}&=\left(\frac{2\pi}{3},\frac{2\sqrt{3}\pi}{3}\right) \equiv \bvec{K}
    \\
    \bvec{Q}_\mathrm{s}&=\left(\frac{\pi}{3},\frac{\sqrt{3}\pi}{3}\right)=\bvec{K}/2,
    \label{eq:wavevec}
\end{align}
where $\bvec{K}$ denotes a corner of the hexagonal Brillouin zone of the triangular lattice.
Note that by momentum conservation and spin-rotation symmetry, there exists an allowed coupling between SDW and CDW order parameters in the free energy of the system. It takes the form
\begin{equation} \label{eq:int_cdw_sdw}
    \mathcal{F}_\mathrm{int} \sim \lambda \rho_{\bvec{Q}_\mathrm{c}}^{\dagger} \left( \vec{S}_{\bvec{Q}_\mathrm{s}} \cdot  \vec{S}_{\bvec{Q}_\mathrm{s}} \right) + \hc,
\end{equation}
where $\rho_{\bvec{Q}_\mathrm{c}}^{\vphantom{\dagger}} = \rho_{-\bvec{Q}_\mathrm{c}}^\dagger$ and similarly $S^z_{\bvec{Q}_\mathrm{s}}$ are the Fourier components of $\rho(\bvec{r})$ and $S^z(\bvec{r})$ at the respective ordering wavevectors, and $\lambda$ is an undetermined coupling constant.

From the interaction $\mathcal{F}_\mathrm{int}$ we may infer that (1) a dominant SDW instability of the system induces a concomitant CDW ordering, (2) the interaction is maximized if the SDW order is collinear, i.e. $\vec{S}_{\bvec{Q}_c}$ is real (up to a global $\Uone$ phase factor), and (3) in such a scenario, the magnitude of the CDW order parameter scales with the square of the SDW order parameter $|\rho_{\bvec{Q}_\mathrm{c}}| \sim |\vec{S}_{\bvec{Q}_\mathrm{s}}|^2$.
One many thus refer to the CDW order as a ``secondary order parameter''.

Ref.~\onlinecite{davoudi08}, which has studied the leading instabilities of the Fermi liquid state in the triangular lattice extended Hubbard model at different densities, provides a useful point of comparison. At filling $\bar{n} = 5/4$, the near-circular Fermi surface does not possess any nesting instability and thus SDW/CDW instabilities do not occur at weak-coupling, but rather require some intermediate coupling strength for which controlled results are rare. However, 
the non-interacting density susceptibility was found to be peaked near $\bvec{K}/2$, and the interacting susceptibility (computed in a two-particle self-consistent approach) in the spin channel suggests an incommensurate spin-density wave order below $\bar{n} \lesssim 1.5$. Note that an exact quantitative comparison is not available as the calculation of Ref.~\onlinecite{davoudi08} applies to the case $U = 4t, V= 1.5t$.  A typical effect, which may be enhanced by the finite cylinder width, is locking of incommensurate order to the lattice, and so it is plausible to regard $\bvec{Q}_\mathrm{s} = \bvec{K}/2$ and $\bvec{Q}_\mathrm{c} = \bvec{K}$ as arising from such a locking.

A somewhat subtle point is the presence of a single particle and charge gap, observed numerically. Insulating states must satisfy \emph{filling constraints}, exemplified by Luttinger's theorem.  Such filling constraints can and should be applied including the effects of spontaneous symmetry breaking.  First let us take a two dimensional point of view, including the observed SDW order.  A featurely electronically trivial state (i.e. one lacking topological order and having a gap) can occur only when the unit cell \emph{of the ordered state} contains an integer number of electrons of each spin (for the collinear order discussed here).  The unit cell in Fig.~\ref{fig:R54Vp0}(b) of the spin pattern contains 6 sites.  The total number of electrons at this filling in this unit cell is $6\times \frac54 = \frac{15}{2}$ which is non-integer, and the number per spin is $\frac{15}{4}$.  This does not satisfy the filling constraint.  In simple terms, there is no possible description of an effective band insulator, using mean-field bands which have been folded by the SDW/CDW order, in which the folded bands are all completely filled or empty. 
In two dimensions, it is expected that the presence of a gap in this situation can occur only in the presence of topological order.
We find this an unlikely scenario.

\begin{figure}
	\centering
	\includegraphics[width=\linewidth]{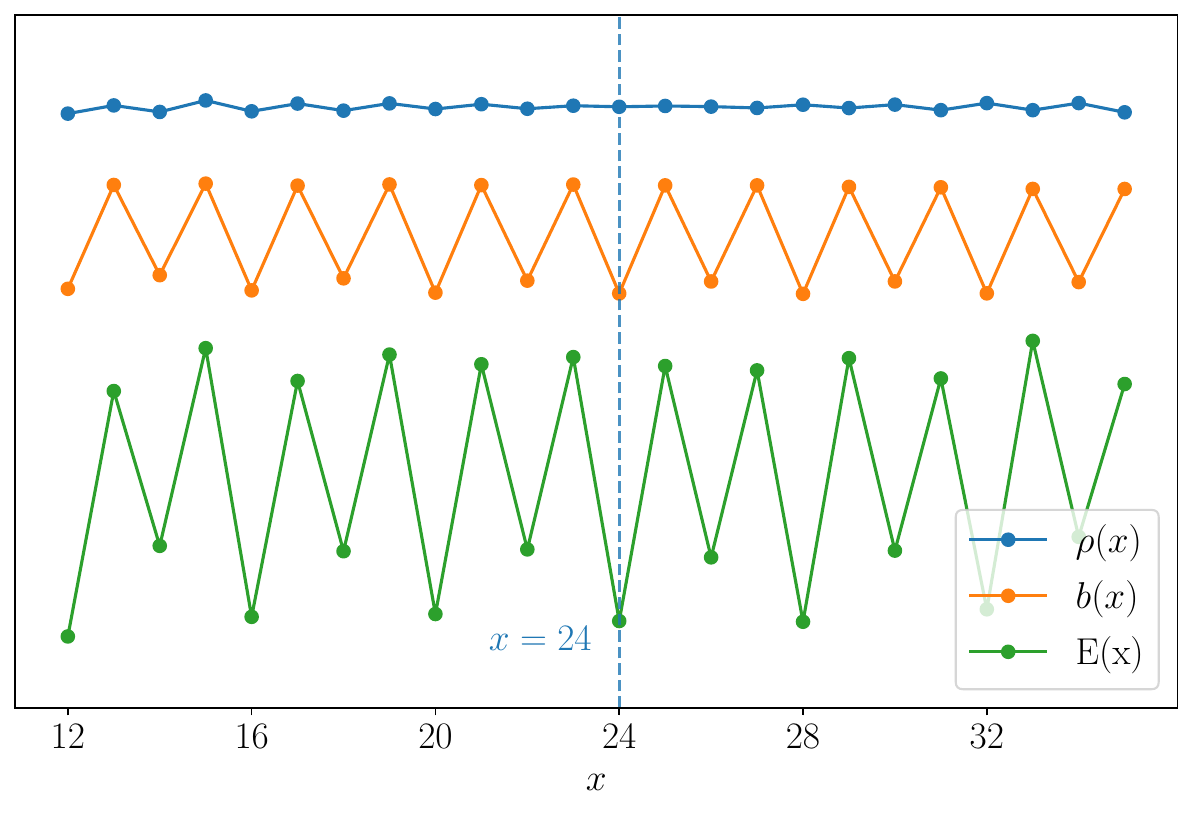}
	\caption{Rung-averaged charge density $\rho(x)$, bond kinetic energy $b(x)$, and entanglement entropy $E(x)$ of the bipartition cutting horizontal bonds at $x$ (see text for definitions), measured by $\SUtwo$ DMRG, with the bond dimension extrapolated to infinity. Different curves are scaled and offset to display the oscillation structure only and the absolute values are meaningless. 
    The center bond of the simulated cylinder is at $x=24.5$.
    The oscillation of the density $\rho(x)$ is negligible compared to the significant oscillations of the bond kinetic energy and entanglement entropy.}
    \label{fig:R54Vp0_EE}
\end{figure}

Instead, since the simulation has a finite cylinder width, we may seek an origin in one-dimensionality.  For a fixed cylinder width, an infinitely long cylinder is a one-dimensional system and must satisfy associated filling constraints, which are known from the Lieb-Schultz-Mattis (LSM) theorem \cite{lieb61,oshikawa00}. 
In a strictly one-dimensional system, there is a subtlety that the SDW order cannot be long-ranged: an order parameter which does not commute with the Hamiltonian and which breaks a continuous symmetry cannot order in one dimension even at zero temperature. 
Indeed, a careful simulation using SU(2) DMRG shows that the SDW order disappears for very large bond dimension, while the CDW order, which does not break a continuous symmetry, remains.  The interpretation is that the two dimensional system has SDW order which however fluctuates quantumly in the long cylinder limit.  The result of this fluctuation can be some additional symmetry breaking.
To this point, we examined in addition to the rung-averaged site density $\rho(x)$, the rung-averaged bond kinetic energy $b(x)$:
\begin{eqnarray}
\rho(x) & = &\frac{1}{L_y}\sum_{y,\sigma}\langle n_\sigma(x,y)\rangle, \nonumber \\
b(x) & = & \frac{1}{L_y}\sum_{y,\sigma}\langle c_{\sigma}^{\dagger}(x,y)c_{\sigma}^{\vphantom{\dagger}}(x+1,y)+\hc \rangle    
\end{eqnarray}
and the entanglement entropy of a bipartition cutting a column of horizontal bonds at position $x+0.5$, denoted as $E(x)$. Here, $x$ and $y$ are \emph{not} cartesian coordinates, but rather correspond to the coordinates $x,y$ along the $\hat{x}$ and $\hat{y}$ directions of a site at position $\bvec{r} = x \hat{x} + y \hat{y}$.
The results are shown in Fig.~\ref{fig:R54Vp0_EE}. 
In the na\"ive SDW/CDW state postulated, this would be uniform, but we observe a two-fold modulation along the $\hat{x}$ direction.
This is the same sort of dimerization which is observed in the gapped phase of one-dimensional spin-1/2 $J_1$-$J_2$ Heisenberg chain.
The additional doubling of the unit cell by this dimerization is sufficient to satisfy the LSM constraints.
We provide a bosonization analysis which rationalizes the appearance of such a dimerization in Appendix~\ref{app:bosonization}.

The conclusion of this somewhat lengthy discussion is that the opening of a single-particle and charge gap in this SDW/CDW state is likely an artifact of the finite cylinder width.  We hypothesize that the system displays a weakly metallic state in the two-dimensional limit.  

\section{Quantum-disordered States} \label{Sec:disorder}
We now turn to the two parameter sets where our DMRG simulation results show no bulk magnetic order.
We consider the system at $\bar{n}=5/3$ (and $V'=0$), where we observe the emergence of a generalized Wigner crystal with singly-occupied sites forming a triangular sublattice of local moments.
Further, we study the case of $\bar{n}= 5/4$, where longer-ranged repulsive interactions, i.e.~$V'=V/\sqrt{3}$, stabilize a generalized Wigner crystal with singly occupied sites forming an effective Kagom\'e lattice, where we suggest that a quantum spin liquid phase may be present.

\subsection{Bulk magnetic disorder at $\bar{n}=5/3, V'=0$}
\label{Sec:R53}
\begin{figure}[ht]
	\centering
	\includegraphics[width=0.9\linewidth]{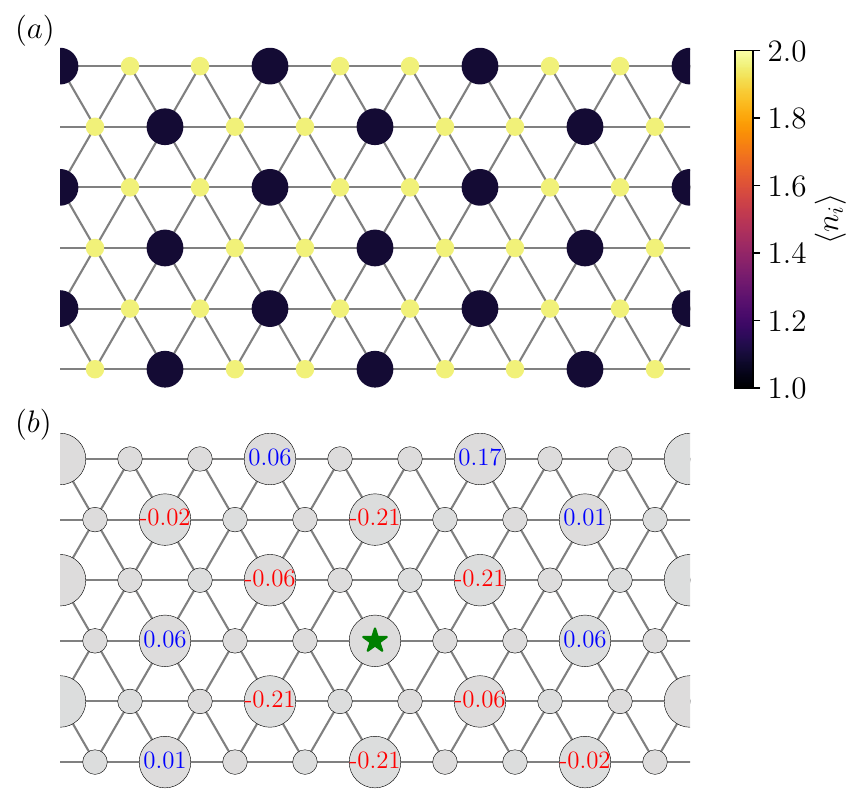}
	\caption{Charge density and spin correlations of the magnetically-disordered state at filling $\bar{n} = 5/3$ with $V'=0$. (a) Illustration of the $\sqrt{3}\times\sqrt{3}$ charge order, giving rise to a triangular lattice of local moments (b) Static spin-spin correlation functions in the magnetically-disordered state (i.e. $\langle \vec{S}_i \rangle =0$ on all sites). The annotated numbers indicate the $\langle\vec{S}_0\cdot\vec{S}_i\rangle$ correlation with the reference site marked by a green star.
    Only the central region of the $N=48 \times 6$ lattice is shown to highlight the bulk phenomenology of the state.
    Here, $U=12t$, $V=U/4$ and $V^{\prime}=0$ with $\Uone$ DMRG and bond dimension $m=25\,000$. \label{fig:R53}
 }
\end{figure}

At filling $\bar{n}=5/3$, there is a strong tendency to form an insulating charge order even at a relatively small $U/t$, as shown in Fig.~\ref{fig:phase diagram}. 
Classical and mean-field calculations \cite{sszb23} predict a Wigner crystallization of a three-sublattice structure ($\sqrt{3} \times \sqrt{3}$ unit cell), where one sublattice site is singly occupied and the other two sublattice sites are occupied by doublons.
Our DMRG simulation confirms that this charge ordering is stable to quantum fluctuations at finite $t/U> 0$, as shown in Fig.~\ref{fig:R53}.
Thus, local moments are expected to emerge at the singly occupied sites, forming an effective triangular lattice with an enlarged lattice constant of $\sqrt{3} a$ (with $a\equiv 1$ denoting the lattice constant of the underlying triangular lattice).
In principle, an effective spin model for exchange interactions between these local moments may be derived -- the first non-trivial interaction corresponds to a nearest-neighbor (on the enlarged triangular lattice of local moments) Heisenberg term, which emerges at fourth order in $t/U$. Note that next-nearest neighbor interactions between these local moments are expected to emerge only at sixth order in $t/U$.
We further note that the cylindrical geometry of our model (with periodic boundary conditions along the $(1/2,\sqrt{3}/2)$ direction with a width of $L_y = 6$) implies that this effective Heisenberg spin model is defined on an effective YC3 geometry (i.e.~there are bonds of the triangular sublattice parallel to the \emph{Cartesian} $y$-axis (i.e. $(0,1)$), and the cylinder is 3 bonds wide).

There have been plenty of numerical studies of Heisenberg or the half-filled Hubbard model on a triangular lattice.
Typically, on YC3 geometries, the ground state has been found to be magnetically ordered: 
Ref.~\onlinecite{saad15} finds that the ground state of the $J_1$-$J_2$ Heisenberg model on a YC3 cylinder geometry has 120$^\circ$ AFM/columnar/dimerized order, depending on the ratio of $J_2/J_1$.
Similarly, Zhu and White \cite{zhu15} find that on YC-odd cylinders, such a $J_1$-$J_2$ Heisenberg model tends to have a dimerized ground state. Studies of the triangular lattice Heisenberg model on XC3 geometries also find dimerization in the ground state \cite{chen13,peng21}. 
Turning toward the half-filled Hubbard model on the triangular lattice, at intermediate $U/t$ a pronounced peak of magnetic structure factor at $\bvec{q}=\bvec{M}$ was found \cite{szasz20}, and an absence of chiral ordering was confirmed \cite{wietek21}. 

We note that our results do not fully agree with any of these previous studies of the effective spin model on 3-leg cylinders: we observe vanishing local spin polarizations and further the absence of any dimer order, chiral order, charge current order and superconductivity pairing order. 
In Fig.~\ref{fig:R53}(b), we annotate the spin correlations $\langle \vec{S}_0\cdot \vec{S}_i\rangle$ of the half-filled sites with the reference site in the center (marked by a green star).
The nearest neighbor spin-spin correlation is antiferromagnetic, and the next-nearest neighbor correlators are ferromagnetic, resembling the antiferromagnetic (ferromagnetic) correlations between nearest (next-nearest) neighbors in a 120$^\circ$ structure stabilized by dominant nearest-neighbor Heisenberg interactions on the triangular lattice (however, we emphasize that no static magnetic ordering is observed).
The spatial anisotropy of the spin-spin correlators visible in Fig.~\ref{fig:R53}(b) can be explained by the absence of an exact $C_3$ symmetry on cylindrical geometries.
The discrepancy with the mentioned studies might be attributed to additional interactions generated by charge fluctuations involving doubly occupied sites. These will be underestimated when truncating a strong-coupling perturbation theory calculation of an effective spin model at low orders in $t/U$ to obtain an effective Heisenberg model.
We also remark that the small width of the effective YC3 cylinder implies that one-dimensional effects may take on a more pronounced role, and further that the effective triangular lattice of local moments has a different termination pattern at the open boundaries than the YC3 cylinder studied previously in the literature. Thus, it is an open question to what extent the observed quantum paramagnetic behaviour is an artifact of the effective one-dimensionality of the system or also persists as a bulk phenomenon.
Determining its precise character and underlying mechanism requires further investigations that lie beyond the scope of the work at hand.

\subsection{Quantum spin liquid at $\bar{n}=5/4, V'=V/\sqrt{3}$} \label{sec:kagome-crystal}
\begin{figure}[ht]
	\centering
	\includegraphics[width=0.9\linewidth]{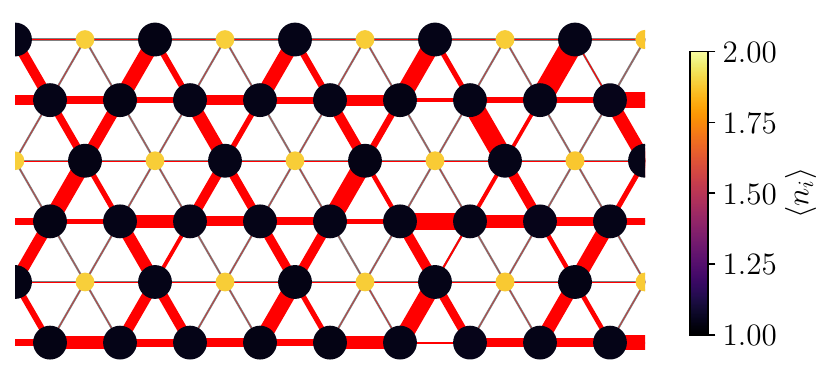}
	\caption{
    Magnetically-disordered state at filling $\bar{n} = 5/4$ with $V'=V/\sqrt{3}$. We observe the absence of any spin polarization $\langle \vec{S}_i \rangle =0$. We encode the nearest-neighbor spin-spin correlation $\braket{\vec{S}_i\cdot\vec{S}_j}$ in the widths of the bonds drawn. 
    The spin-spin correlators involving a doubly occupied site are negligible.
    We only show the middle section of the $48 \times 6$ cyclinder to emphasize bulk phenomenlogy.
    Here, we show results for $U=12t$, $V=U/4$ and $V^{\prime}=V/\sqrt{3}$ from $\SUtwo$ DMRG simulations with bond dimension $m=20\,000$ (equivalent to a bond dimension of $m \sim 57\,000$ in $\Uone$ DMRG simulations).
 \label{fig:R54Vp}}
\end{figure}

\begin{figure}[t]
    \centering
    \includegraphics[width=\columnwidth]{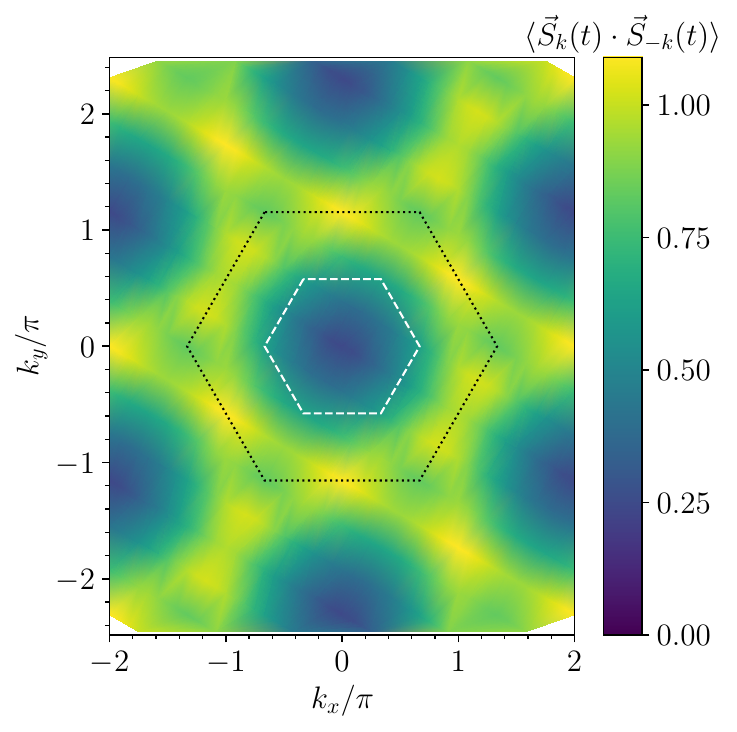}
    \caption{Equal-time structure factor $\langle S_{k}(t) \cdot S_{-k}(t) \rangle$ in momentum space.
    The dotted black (dashed white) line indicates the Brillouin zone of the parent triangular (emergent Kagom\'e) lattice. 
    \label{fig:sfactQSL}}
\end{figure}

As discussed in Sec.~\ref{sec:R54_Vp0}, at filling $\bar{n}=5/4$, a finite nearest-neighbor repulsion $V>0$ is not sufficient to stabilize a generalized Wigner crystal, but rather only induces a weak charge-density wave modulation concomitant with a spin-density wave instability.
However, classical and mean-field arguments indicate that switching on a second-nearest neighbor repulsive interaction $V'=V/\sqrt{3}$ stabilizes Kagom\'e charge crystal where half-filled sites form an effective Kagom\'e lattice.
Our DMRG simulations at finite $t/U$ support this scenario: a stable Kagom\'e charge crystal can be seen in the site-resolved occupation numbers displayed in Fig.~\ref{fig:R54Vp}.
We find that our $\SUtwo$-DMRG calculations which enforce the full $\SUtwo$ spin rotation symmetry of the Hamiltonian yield a lower ground state energy than the $\Uone$-enforcing DMRG (for details, see Appendix.~\ref{sec:U1vsSU2}).
We conclude that the ground state of the model does not feature spontaneous spin-rotation symmetry breaking, and rather corresponds to a paramagnetic (quantum-disordered) state.

To characterize this quantum-disordered magnetic state, we display the nearest-neighbor spin-spin correlators $\langle\vec{S}_i\cdot\vec{S}_j\rangle$ of this state in Fig.~\ref{fig:R54Vp} (note that correlators involving doubly occupied sites are negligible).
We observe that the correlators are mostly uniform throughout the lattice, with small anisotropic corrections that may be attributed to boundary and finite-size effects.
A systematic study of these effects on different system sizes and geometries is left as a task for further study, but our results do not appear to suggest any spontaneous breaking of the emergent Kagom\'e lattice's rotational or translational symmetries which would necessarily be the case for dimerized states or possibly nematic spin-liquid behaviour. Further, we find no signatures of potential chiral ordering in the system at hand.
In Fig.~\ref{fig:sfactQSL}, we show the equal-time structure factor $\braket{\vec{S}_k(t)\cdot \vec{S}_{-k}(t)}$, corresponding to the integrated dynamical structure factor in frequency space.
Broad peaks near the boundary of the Brillouin zone can be observed, as also found in previous studies of frustrated Kagom\'e quantum magnets \cite{messio10,iqbal13,punk14,Fu2015,kolley15,halimeh16,messio17,chern21}, further supporting the absence of magnetic ordering.
We therefore conjecture that the ground state of the model in the parameter regime considered is given by a nonchiral lattice-symmetry preserving quantum spin liquid.

We have also computed spin-spin, single-particle, superconducting pair-pair, and dimer correlators as a function of distance, which are found to be short ranged and exhibit an exponential decay, as shown in Fig.~\ref{fig:combined corr}b.
While it thus may be tempting to speculate that the spin liquid state at hand is gapped, we caution the reader that the apparently gapped behaviour may also be an artifact: gapless $\Uone$ Dirac spin liquids can lead to gapped ground states on cylinder geometries if the lattice size/boundary conditions are not commensurate with the gapless points in the spectrum \cite{he17,hu19}.
It is unclear if the spin-liquid state of local moments found in our system is in the same phase as putative quantum spin liquid states in Kagom\'e lattice Heisenberg models. Their exact nature has remained elusive and subject to debates:
mean field and variational Monte Carlo calculations predict a $\Uone$ Dirac (algebraic) spin liquid as the ground state \cite{hermele08,iqbal13,iqbal14,iqbal15} which is supported by several DMRG studies utilizing flux insertion \cite{he17,zhu18}.
Notably, Neutron scattering experiments of candidate materials have reported the absence of a spin gap \cite{han12,fak12}. 
On the other hand, DMRG studies have also suggested a gapped $\Ztwo$ spin liquid state \cite{jiang08,yan11,jiang12,depenbrock12}, and one-loop calculations of the dynamical structure factor of $\Ztwo$ spin liquid with vison excitations find qualitative agreement with Neutron scattering experiments on Herbertsmithites \cite{punk14,Fu2015}.
Deforming the nearest-neighbor Heisenberg model on the Kagom\'e lattice, e.g. by second- and third-nearest neighbor interactions or three-spin terms, may stabilize a chiral spin liquid ground state \cite{messio12,gong14,bauer14,wietek15,hu15}, which however is inconsistent with the time-reversal symmetry-preserving state found in our simulations.

\section{Summary and conclusion} \label{Sec:Summary}

Motivated by the observation of Mott-insulating behaviour and generalized Wigner crystals in moiré TMD, we have performed extensive DMRG simulations for an extended Hubbard model on the triangular lattice at certain fractional fillings, with the goal of elucidating possible magnetic states that emerge in these Wigner crystal states.

We have presented conclusive evidence that at filling $\bar{n}=4/3$, a generalized Wigner crystal of singly occupied sites forms an emergent honeycomb lattice for sufficiently large $U,V \gg t$.
Exchange interactions lead to a zigzag antiferromagnetic ordering of the associated local moments, as also captured by an effective Heisenberg spin model derived in strong-coupling perturbation theory.
At filling $\bar{n}=5/3$, we observe a generalized Wigner crystal where singly occupied sites form an emergent triangular lattice with lattice constant $\sqrt{3}$. Our simulations do not provide evidence for magnetic or nematic ordering, indicative of a quantum-disordered state. While such state may be connected to a spin liquid phase in bulk (2D) systems, we caution that there may be significant finite-size effects, owing to the small width of the emergent triangular lattice and call for systematic numerical investigations.

Turning to $\bar{n}=5/4$, we find that no Wigner crystal is stabilized if only nearest-neighbor repulsion is present. Instead, we find that there is an intermediate-coupling instability of the metallic Fermi surface towards a spin-density wave order with concomitant \emph{weak} CDW ordering.
Upon switching on a second-nearest neighbor interaction $V'=V/\sqrt{3}$, corresponding to longer-ranged interactions (e.g. by reducing screening by an adjacent gating layer), a Kagom\'e charge crystal is stabilized, where singly occupied sites form an emergent Kagom\'e lattice. Local moments interacting with antiferromagnetic exchange interactions on a Kagom\'e lattice are a paradigmatic example for a strongly frustrated spin system.
We find that, in the context of the Hubbard model that we study, the local moments in this Kagom\'e charge crystal are fully disordered, and observe short-ranged spin-spin, bond-bond, single particle, chiral correlations.

Taken together, these observations are consistent with a time reversal-symmetric spin-liquid ground state at $\bar{n} = 5/4$ and finite $V'$.
Further systematic studies on different system sizes and geometries are required to pin down the nature of this putative spin liquid state, and to what extent it can be connected to the much-debated spin-liquid ground state of the Kagom\'e lattice Heisenberg model.

While we have focused on filling factors $\bar{n} > 1$, we expected that qualitatively similar results may also emerge for rational fillings $\bar{n} < 1$ \cite{zhou2024quantum,biborski24}. An interesting direction for further study consists in weakly doping away from rational fillings: while in some cases, adding a weak density of carriers promotes ferromagnetism \cite{ufps24,morera23,zhao24,ciorciaro23}, on some lattice geometries antiferromagnetic correlations may become enhanced \cite{haerter05,sposetti14,morera23}.
Another intriguing direction could be the moir\'e Hubbard model combined with a magnetic or layer-pseudo magnetic field, i.e.~a Hubbard-Hofstadter model, where the chiral spin liquid state can serve as a promising candidate in the intermediate interaction regime\cite{Kuhlenkamp24,divic24}.

Overall, our results show that the Hubbard model with extended repulsive interactions at fractional fillings realizes a rich structure of magnetically ordered and disordered states.
These can be accessed in a targeted manner by controlling the system's filling, whereupon stable generalized Wigner crystals emerge which constitute (possibly highly frustrated) lattices of interacting local moments.

An open question pertains to the experimental detection of antiferromagnetically ordered and putative quantum spin-liquid states in moiré TMD heterostructures. Neutron scattering, conventionally used for atomic magnets, appears inapplicable due to sample size limitations. Optical methods have been successfully used to resolve time reversal-symmetry breaking ferromagnetism in moiré TMD \cite{anderson23,ciorciaro23,zhao24}, but provide limited insight into dynamical and finite-momentum correlations. Here, advanced tunneling spectroscopy set-ups may constitute a promising avenue for further theoretical and experimental study \cite{koenig20,peri24,pichler24}.


\begin{acknowledgments}
We thank Yu-Ping Lin and Jia-Xin Zhang for helpful conversations.
U.F.P.S. acknowledges support from the Deutsche Forschungsgemeinschaft (DFG, German Research Foundation) via SFB SFB 1238, Project ID No. 277146847, the Emmy Noether Program, Project ID No. 544397233 and a Walter Benjamin fellowship, Project ID No. 449890867.
This research was supported in part by the National Science Foundation under Grant No. NSF PHY-1748958. H.-C.J. was supported by the Department of Energy (DOE), Office of Sciences, Basic Energy Sciences, Materials Sciences and Engineering Division, under Contract No. DE-AC02-76SF00515.  L.B. and Z.S were supported by the NSF CMMT program under Grant No. DMR-2419871, and by the Simons Collaboration on Ultra-Quantum Matter, which is a grant from the Simons Foundation (Grant No. 651440).  
\end{acknowledgments}


\bibliography{hubbard_dmrg_bib}

\begin{thebibliography}{78}%
\makeatletter
\providecommand \@ifxundefined [1]{%
 \@ifx{#1\undefined}
}%
\providecommand \@ifnum [1]{%
 \ifnum #1\expandafter \@firstoftwo
 \else \expandafter \@secondoftwo
 \fi
}%
\providecommand \@ifx [1]{%
 \ifx #1\expandafter \@firstoftwo
 \else \expandafter \@secondoftwo
 \fi
}%
\providecommand \natexlab [1]{#1}%
\providecommand \enquote  [1]{``#1''}%
\providecommand \bibnamefont  [1]{#1}%
\providecommand \bibfnamefont [1]{#1}%
\providecommand \citenamefont [1]{#1}%
\providecommand \href@noop [0]{\@secondoftwo}%
\providecommand \href [0]{\begingroup \@sanitize@url \@href}%
\providecommand \@href[1]{\@@startlink{#1}\@@href}%
\providecommand \@@href[1]{\endgroup#1\@@endlink}%
\providecommand \@sanitize@url [0]{\catcode `\\12\catcode `\$12\catcode `\&12\catcode `\#12\catcode `\^12\catcode `\_12\catcode `\%12\relax}%
\providecommand \@@startlink[1]{}%
\providecommand \@@endlink[0]{}%
\providecommand \url  [0]{\begingroup\@sanitize@url \@url }%
\providecommand \@url [1]{\endgroup\@href {#1}{\urlprefix }}%
\providecommand \urlprefix  [0]{URL }%
\providecommand \Eprint [0]{\href }%
\providecommand \doibase [0]{https://doi.org/}%
\providecommand \selectlanguage [0]{\@gobble}%
\providecommand \bibinfo  [0]{\@secondoftwo}%
\providecommand \bibfield  [0]{\@secondoftwo}%
\providecommand \translation [1]{[#1]}%
\providecommand \BibitemOpen [0]{}%
\providecommand \bibitemStop [0]{}%
\providecommand \bibitemNoStop [0]{.\EOS\space}%
\providecommand \EOS [0]{\spacefactor3000\relax}%
\providecommand \BibitemShut  [1]{\csname bibitem#1\endcsname}%
\let\auto@bib@innerbib\@empty
\bibitem [{\citenamefont {Damascelli}\ \emph {et~al.}(2003)\citenamefont {Damascelli}, \citenamefont {Hussain},\ and\ \citenamefont {Shen}}]{damascelli03}%
  \BibitemOpen
  \bibfield  {author} {\bibinfo {author} {\bibfnamefont {A.}~\bibnamefont {Damascelli}}, \bibinfo {author} {\bibfnamefont {Z.}~\bibnamefont {Hussain}},\ and\ \bibinfo {author} {\bibfnamefont {Z.-X.}\ \bibnamefont {Shen}},\ }\href {https://doi.org/10.1103/RevModPhys.75.473} {\bibfield  {journal} {\bibinfo  {journal} {Rev. Mod. Phys.}\ }\textbf {\bibinfo {volume} {75}},\ \bibinfo {pages} {473} (\bibinfo {year} {2003})}\BibitemShut {NoStop}%
\bibitem [{\citenamefont {Lee}\ \emph {et~al.}(2006)\citenamefont {Lee}, \citenamefont {Nagaosa},\ and\ \citenamefont {Wen}}]{lee06}%
  \BibitemOpen
  \bibfield  {author} {\bibinfo {author} {\bibfnamefont {P.~A.}\ \bibnamefont {Lee}}, \bibinfo {author} {\bibfnamefont {N.}~\bibnamefont {Nagaosa}},\ and\ \bibinfo {author} {\bibfnamefont {X.-G.}\ \bibnamefont {Wen}},\ }\href {https://doi.org/10.1103/RevModPhys.78.17} {\bibfield  {journal} {\bibinfo  {journal} {Rev. Mod. Phys.}\ }\textbf {\bibinfo {volume} {78}},\ \bibinfo {pages} {17} (\bibinfo {year} {2006})}\BibitemShut {NoStop}%
\bibitem [{\citenamefont {Lee}\ and\ \citenamefont {Lee}(2005)}]{lee05}%
  \BibitemOpen
  \bibfield  {author} {\bibinfo {author} {\bibfnamefont {S.-S.}\ \bibnamefont {Lee}}\ and\ \bibinfo {author} {\bibfnamefont {P.~A.}\ \bibnamefont {Lee}},\ }\href {https://doi.org/10.1103/PhysRevLett.95.036403} {\bibfield  {journal} {\bibinfo  {journal} {Phys. Rev. Lett.}\ }\textbf {\bibinfo {volume} {95}},\ \bibinfo {pages} {036403} (\bibinfo {year} {2005})}\BibitemShut {NoStop}%
\bibitem [{\citenamefont {Savary}\ and\ \citenamefont {Balents}(2016)}]{savary17}%
  \BibitemOpen
  \bibfield  {author} {\bibinfo {author} {\bibfnamefont {L.}~\bibnamefont {Savary}}\ and\ \bibinfo {author} {\bibfnamefont {L.}~\bibnamefont {Balents}},\ }\href {https://doi.org/10.1088/0034-4885/80/1/016502} {\bibfield  {journal} {\bibinfo  {journal} {Reports on Progress in Physics}\ }\textbf {\bibinfo {volume} {80}},\ \bibinfo {pages} {016502} (\bibinfo {year} {2016})}\BibitemShut {NoStop}%
\bibitem [{\citenamefont {Arovas}\ \emph {et~al.}(2022)\citenamefont {Arovas}, \citenamefont {Berg}, \citenamefont {Kivelson},\ and\ \citenamefont {Raghu}}]{arovas21}%
  \BibitemOpen
  \bibfield  {author} {\bibinfo {author} {\bibfnamefont {D.~P.}\ \bibnamefont {Arovas}}, \bibinfo {author} {\bibfnamefont {E.}~\bibnamefont {Berg}}, \bibinfo {author} {\bibfnamefont {S.~A.}\ \bibnamefont {Kivelson}},\ and\ \bibinfo {author} {\bibfnamefont {S.}~\bibnamefont {Raghu}},\ }\href {https://doi.org/https://doi.org/10.1146/annurev-conmatphys-031620-102024} {\bibfield  {journal} {\bibinfo  {journal} {Annual Review of Condensed Matter Physics}\ }\textbf {\bibinfo {volume} {13}},\ \bibinfo {pages} {239} (\bibinfo {year} {2022})}\BibitemShut {NoStop}%
\bibitem [{\citenamefont {Jiang}\ \emph {et~al.}(2020)\citenamefont {Jiang}, \citenamefont {Zaanen}, \citenamefont {Devereaux},\ and\ \citenamefont {Jiang}}]{jiang20}%
  \BibitemOpen
  \bibfield  {author} {\bibinfo {author} {\bibfnamefont {Y.-F.}\ \bibnamefont {Jiang}}, \bibinfo {author} {\bibfnamefont {J.}~\bibnamefont {Zaanen}}, \bibinfo {author} {\bibfnamefont {T.~P.}\ \bibnamefont {Devereaux}},\ and\ \bibinfo {author} {\bibfnamefont {H.-C.}\ \bibnamefont {Jiang}},\ }\href {https://doi.org/10.1103/PhysRevResearch.2.033073} {\bibfield  {journal} {\bibinfo  {journal} {Phys. Rev. Res.}\ }\textbf {\bibinfo {volume} {2}},\ \bibinfo {pages} {033073} (\bibinfo {year} {2020})}\BibitemShut {NoStop}%
\bibitem [{\citenamefont {Zhang}\ and\ \citenamefont {Callaway}(1989)}]{zhang1989extended}%
  \BibitemOpen
  \bibfield  {author} {\bibinfo {author} {\bibfnamefont {Y.}~\bibnamefont {Zhang}}\ and\ \bibinfo {author} {\bibfnamefont {J.}~\bibnamefont {Callaway}},\ }\href@noop {} {\bibfield  {journal} {\bibinfo  {journal} {Physical Review B}\ }\textbf {\bibinfo {volume} {39}},\ \bibinfo {pages} {9397} (\bibinfo {year} {1989})}\BibitemShut {NoStop}%
\bibitem [{\citenamefont {Terletska}\ \emph {et~al.}(2017)\citenamefont {Terletska}, \citenamefont {Chen},\ and\ \citenamefont {Gull}}]{PhysRevB.95.115149}%
  \BibitemOpen
  \bibfield  {author} {\bibinfo {author} {\bibfnamefont {H.}~\bibnamefont {Terletska}}, \bibinfo {author} {\bibfnamefont {T.}~\bibnamefont {Chen}},\ and\ \bibinfo {author} {\bibfnamefont {E.}~\bibnamefont {Gull}},\ }\href {https://doi.org/10.1103/PhysRevB.95.115149} {\bibfield  {journal} {\bibinfo  {journal} {Phys. Rev. B}\ }\textbf {\bibinfo {volume} {95}},\ \bibinfo {pages} {115149} (\bibinfo {year} {2017})}\BibitemShut {NoStop}%
\bibitem [{\citenamefont {Aichhorn}\ \emph {et~al.}(2004)\citenamefont {Aichhorn}, \citenamefont {Evertz}, \citenamefont {von~der Linden},\ and\ \citenamefont {Potthoff}}]{aichhorn2004charge}%
  \BibitemOpen
  \bibfield  {author} {\bibinfo {author} {\bibfnamefont {M.}~\bibnamefont {Aichhorn}}, \bibinfo {author} {\bibfnamefont {H.~G.}\ \bibnamefont {Evertz}}, \bibinfo {author} {\bibfnamefont {W.}~\bibnamefont {von~der Linden}},\ and\ \bibinfo {author} {\bibfnamefont {M.}~\bibnamefont {Potthoff}},\ }\href@noop {} {\bibfield  {journal} {\bibinfo  {journal} {Physical Review B—Condensed Matter and Materials Physics}\ }\textbf {\bibinfo {volume} {70}},\ \bibinfo {pages} {235107} (\bibinfo {year} {2004})}\BibitemShut {NoStop}%
\bibitem [{\citenamefont {Sandvik}\ \emph {et~al.}(2004)\citenamefont {Sandvik}, \citenamefont {Balents},\ and\ \citenamefont {Campbell}}]{sandvik2004ground}%
  \BibitemOpen
  \bibfield  {author} {\bibinfo {author} {\bibfnamefont {A.~W.}\ \bibnamefont {Sandvik}}, \bibinfo {author} {\bibfnamefont {L.}~\bibnamefont {Balents}},\ and\ \bibinfo {author} {\bibfnamefont {D.~K.}\ \bibnamefont {Campbell}},\ }\href@noop {} {\bibfield  {journal} {\bibinfo  {journal} {Physical review letters}\ }\textbf {\bibinfo {volume} {92}},\ \bibinfo {pages} {236401} (\bibinfo {year} {2004})}\BibitemShut {NoStop}%
\bibitem [{\citenamefont {Tocchio}\ \emph {et~al.}(2014)\citenamefont {Tocchio}, \citenamefont {Gros}, \citenamefont {Zhang},\ and\ \citenamefont {Eggert}}]{PhysRevLett.113.246405}%
  \BibitemOpen
  \bibfield  {author} {\bibinfo {author} {\bibfnamefont {L.~F.}\ \bibnamefont {Tocchio}}, \bibinfo {author} {\bibfnamefont {C.}~\bibnamefont {Gros}}, \bibinfo {author} {\bibfnamefont {X.-F.}\ \bibnamefont {Zhang}},\ and\ \bibinfo {author} {\bibfnamefont {S.}~\bibnamefont {Eggert}},\ }\href {https://doi.org/10.1103/PhysRevLett.113.246405} {\bibfield  {journal} {\bibinfo  {journal} {Phys. Rev. Lett.}\ }\textbf {\bibinfo {volume} {113}},\ \bibinfo {pages} {246405} (\bibinfo {year} {2014})}\BibitemShut {NoStop}%
\bibitem [{\citenamefont {Wu}\ \emph {et~al.}(2018)\citenamefont {Wu}, \citenamefont {Lovorn}, \citenamefont {Tutuc},\ and\ \citenamefont {MacDonald}}]{wu18}%
  \BibitemOpen
  \bibfield  {author} {\bibinfo {author} {\bibfnamefont {F.}~\bibnamefont {Wu}}, \bibinfo {author} {\bibfnamefont {T.}~\bibnamefont {Lovorn}}, \bibinfo {author} {\bibfnamefont {E.}~\bibnamefont {Tutuc}},\ and\ \bibinfo {author} {\bibfnamefont {A.~H.}\ \bibnamefont {MacDonald}},\ }\href@noop {} {\bibfield  {journal} {\bibinfo  {journal} {Phys. Rev. Lett.}\ }\textbf {\bibinfo {volume} {121}},\ \bibinfo {pages} {026402} (\bibinfo {year} {2018})}\BibitemShut {NoStop}%
\bibitem [{\citenamefont {Pan}\ \emph {et~al.}(2020{\natexlab{a}})\citenamefont {Pan}, \citenamefont {Wu},\ and\ \citenamefont {Das~Sarma}}]{pan20a}%
  \BibitemOpen
  \bibfield  {author} {\bibinfo {author} {\bibfnamefont {H.}~\bibnamefont {Pan}}, \bibinfo {author} {\bibfnamefont {F.}~\bibnamefont {Wu}},\ and\ \bibinfo {author} {\bibfnamefont {S.}~\bibnamefont {Das~Sarma}},\ }\href {https://link.aps.org/doi/10.1103/PhysRevResearch.2.033087} {\bibfield  {journal} {\bibinfo  {journal} {Phys. Rev. Res.}\ }\textbf {\bibinfo {volume} {2}},\ \bibinfo {pages} {033087} (\bibinfo {year} {2020}{\natexlab{a}})}\BibitemShut {NoStop}%
\bibitem [{\citenamefont {Wang}\ \emph {et~al.}(2020)\citenamefont {Wang}, \citenamefont {Shih}, \citenamefont {Ghiotto}, \citenamefont {Xian}, \citenamefont {Rhodes}, \citenamefont {Tan}, \citenamefont {Claassen}, \citenamefont {Kennes}, \citenamefont {Bai}, \citenamefont {Kim}, \citenamefont {Watanabe}, \citenamefont {Taniguchi}, \citenamefont {Zhu}, \citenamefont {Hone}, \citenamefont {Rubio}, \citenamefont {Pasupathy},\ and\ \citenamefont {Dean}}]{wang20}%
  \BibitemOpen
  \bibfield  {author} {\bibinfo {author} {\bibfnamefont {L.}~\bibnamefont {Wang}}, \bibinfo {author} {\bibfnamefont {E.-M.}\ \bibnamefont {Shih}}, \bibinfo {author} {\bibfnamefont {A.}~\bibnamefont {Ghiotto}}, \bibinfo {author} {\bibfnamefont {L.}~\bibnamefont {Xian}}, \bibinfo {author} {\bibfnamefont {D.~A.}\ \bibnamefont {Rhodes}}, \bibinfo {author} {\bibfnamefont {C.}~\bibnamefont {Tan}}, \bibinfo {author} {\bibfnamefont {M.}~\bibnamefont {Claassen}}, \bibinfo {author} {\bibfnamefont {D.~M.}\ \bibnamefont {Kennes}}, \bibinfo {author} {\bibfnamefont {Y.}~\bibnamefont {Bai}}, \bibinfo {author} {\bibfnamefont {B.}~\bibnamefont {Kim}}, \bibinfo {author} {\bibfnamefont {K.}~\bibnamefont {Watanabe}}, \bibinfo {author} {\bibfnamefont {T.}~\bibnamefont {Taniguchi}}, \bibinfo {author} {\bibfnamefont {X.}~\bibnamefont {Zhu}}, \bibinfo {author} {\bibfnamefont {J.}~\bibnamefont {Hone}}, \bibinfo {author} {\bibfnamefont {A.}~\bibnamefont {Rubio}}, \bibinfo {author} {\bibfnamefont {A.~N.}\ \bibnamefont {Pasupathy}},\
  and\ \bibinfo {author} {\bibfnamefont {C.~R.}\ \bibnamefont {Dean}},\ }\href {https://doi.org/10.1038/s41563-020-0708-6} {\bibfield  {journal} {\bibinfo  {journal} {Nature Materials}\ }\textbf {\bibinfo {volume} {19}},\ \bibinfo {pages} {861} (\bibinfo {year} {2020})}\BibitemShut {NoStop}%
\bibitem [{\citenamefont {Li}\ \emph {et~al.}(2021{\natexlab{a}})\citenamefont {Li}, \citenamefont {Jiang}, \citenamefont {Li}, \citenamefont {Zhang}, \citenamefont {Kang}, \citenamefont {Zhu}, \citenamefont {Watanabe}, \citenamefont {Taniguchi}, \citenamefont {Chowdhury}, \citenamefont {Fu}, \citenamefont {Shan},\ and\ \citenamefont {Mak}}]{limak21}%
  \BibitemOpen
  \bibfield  {author} {\bibinfo {author} {\bibfnamefont {T.}~\bibnamefont {Li}}, \bibinfo {author} {\bibfnamefont {S.}~\bibnamefont {Jiang}}, \bibinfo {author} {\bibfnamefont {L.}~\bibnamefont {Li}}, \bibinfo {author} {\bibfnamefont {Y.}~\bibnamefont {Zhang}}, \bibinfo {author} {\bibfnamefont {K.}~\bibnamefont {Kang}}, \bibinfo {author} {\bibfnamefont {J.}~\bibnamefont {Zhu}}, \bibinfo {author} {\bibfnamefont {K.}~\bibnamefont {Watanabe}}, \bibinfo {author} {\bibfnamefont {T.}~\bibnamefont {Taniguchi}}, \bibinfo {author} {\bibfnamefont {D.}~\bibnamefont {Chowdhury}}, \bibinfo {author} {\bibfnamefont {L.}~\bibnamefont {Fu}}, \bibinfo {author} {\bibfnamefont {J.}~\bibnamefont {Shan}},\ and\ \bibinfo {author} {\bibfnamefont {K.~F.}\ \bibnamefont {Mak}},\ }\href {https://doi.org/10.1038/s41586-021-03853-0} {\bibfield  {journal} {\bibinfo  {journal} {Nature}\ }\textbf {\bibinfo {volume} {597}},\ \bibinfo {pages} {350} (\bibinfo {year} {2021}{\natexlab{a}})}\BibitemShut {NoStop}%
\bibitem [{\citenamefont {Regan}\ \emph {et~al.}(2020)\citenamefont {Regan}, \citenamefont {Wang}, \citenamefont {Jin}, \citenamefont {Bakti~Utama}, \citenamefont {Gao}, \citenamefont {Wei}, \citenamefont {Zhao}, \citenamefont {Zhao}, \citenamefont {Zhang}, \citenamefont {Yumigeta}, \citenamefont {Blei}, \citenamefont {Carlstr{\"o}m}, \citenamefont {Watanabe}, \citenamefont {Taniguchi}, \citenamefont {Tongay}, \citenamefont {Crommie}, \citenamefont {Zettl},\ and\ \citenamefont {Wang}}]{regan20}%
  \BibitemOpen
  \bibfield  {author} {\bibinfo {author} {\bibfnamefont {E.~C.}\ \bibnamefont {Regan}}, \bibinfo {author} {\bibfnamefont {D.}~\bibnamefont {Wang}}, \bibinfo {author} {\bibfnamefont {C.}~\bibnamefont {Jin}}, \bibinfo {author} {\bibfnamefont {M.~I.}\ \bibnamefont {Bakti~Utama}}, \bibinfo {author} {\bibfnamefont {B.}~\bibnamefont {Gao}}, \bibinfo {author} {\bibfnamefont {X.}~\bibnamefont {Wei}}, \bibinfo {author} {\bibfnamefont {S.}~\bibnamefont {Zhao}}, \bibinfo {author} {\bibfnamefont {W.}~\bibnamefont {Zhao}}, \bibinfo {author} {\bibfnamefont {Z.}~\bibnamefont {Zhang}}, \bibinfo {author} {\bibfnamefont {K.}~\bibnamefont {Yumigeta}}, \bibinfo {author} {\bibfnamefont {M.}~\bibnamefont {Blei}}, \bibinfo {author} {\bibfnamefont {J.~D.}\ \bibnamefont {Carlstr{\"o}m}}, \bibinfo {author} {\bibfnamefont {K.}~\bibnamefont {Watanabe}}, \bibinfo {author} {\bibfnamefont {T.}~\bibnamefont {Taniguchi}}, \bibinfo {author} {\bibfnamefont {S.}~\bibnamefont {Tongay}}, \bibinfo {author} {\bibfnamefont {M.}~\bibnamefont
  {Crommie}}, \bibinfo {author} {\bibfnamefont {A.}~\bibnamefont {Zettl}},\ and\ \bibinfo {author} {\bibfnamefont {F.}~\bibnamefont {Wang}},\ }\href {https://doi.org/10.1038/s41586-020-2092-4} {\bibfield  {journal} {\bibinfo  {journal} {Nature}\ }\textbf {\bibinfo {volume} {579}},\ \bibinfo {pages} {359} (\bibinfo {year} {2020})}\BibitemShut {NoStop}%
\bibitem [{\citenamefont {Li}\ \emph {et~al.}(2021{\natexlab{b}})\citenamefont {Li}, \citenamefont {Li}, \citenamefont {Regan}, \citenamefont {Wang}, \citenamefont {Zhao}, \citenamefont {Kahn}, \citenamefont {Yumigeta}, \citenamefont {Blei}, \citenamefont {Taniguchi}, \citenamefont {Watanabe}, \citenamefont {Tongay}, \citenamefont {Zettl}, \citenamefont {Crommie},\ and\ \citenamefont {Wang}}]{liwang21}%
  \BibitemOpen
  \bibfield  {author} {\bibinfo {author} {\bibfnamefont {H.}~\bibnamefont {Li}}, \bibinfo {author} {\bibfnamefont {S.}~\bibnamefont {Li}}, \bibinfo {author} {\bibfnamefont {E.~C.}\ \bibnamefont {Regan}}, \bibinfo {author} {\bibfnamefont {D.}~\bibnamefont {Wang}}, \bibinfo {author} {\bibfnamefont {W.}~\bibnamefont {Zhao}}, \bibinfo {author} {\bibfnamefont {S.}~\bibnamefont {Kahn}}, \bibinfo {author} {\bibfnamefont {K.}~\bibnamefont {Yumigeta}}, \bibinfo {author} {\bibfnamefont {M.}~\bibnamefont {Blei}}, \bibinfo {author} {\bibfnamefont {T.}~\bibnamefont {Taniguchi}}, \bibinfo {author} {\bibfnamefont {K.}~\bibnamefont {Watanabe}}, \bibinfo {author} {\bibfnamefont {S.}~\bibnamefont {Tongay}}, \bibinfo {author} {\bibfnamefont {A.}~\bibnamefont {Zettl}}, \bibinfo {author} {\bibfnamefont {M.~F.}\ \bibnamefont {Crommie}},\ and\ \bibinfo {author} {\bibfnamefont {F.}~\bibnamefont {Wang}},\ }\href {https://doi.org/10.1038/s41586-021-03874-9} {\bibfield  {journal} {\bibinfo  {journal} {Nature}\ }\textbf {\bibinfo
  {volume} {597}},\ \bibinfo {pages} {650} (\bibinfo {year} {2021}{\natexlab{b}})}\BibitemShut {NoStop}%
\bibitem [{\citenamefont {Pan}\ \emph {et~al.}(2020{\natexlab{b}})\citenamefont {Pan}, \citenamefont {Wu},\ and\ \citenamefont {Das~Sarma}}]{pan20b}%
  \BibitemOpen
  \bibfield  {author} {\bibinfo {author} {\bibfnamefont {H.}~\bibnamefont {Pan}}, \bibinfo {author} {\bibfnamefont {F.}~\bibnamefont {Wu}},\ and\ \bibinfo {author} {\bibfnamefont {S.}~\bibnamefont {Das~Sarma}},\ }\href@noop {} {\bibfield  {journal} {\bibinfo  {journal} {Phys. Rev. B}\ }\textbf {\bibinfo {volume} {102}},\ \bibinfo {pages} {201104} (\bibinfo {year} {2020}{\natexlab{b}})}\BibitemShut {NoStop}%
\bibitem [{\citenamefont {Hu}\ and\ \citenamefont {MacDonald}(2021)}]{hu21}%
  \BibitemOpen
  \bibfield  {author} {\bibinfo {author} {\bibfnamefont {N.~C.}\ \bibnamefont {Hu}}\ and\ \bibinfo {author} {\bibfnamefont {A.~H.}\ \bibnamefont {MacDonald}},\ }\href {https://doi.org/10.1103/PhysRevB.104.214403} {\bibfield  {journal} {\bibinfo  {journal} {Phys. Rev. B}\ }\textbf {\bibinfo {volume} {104}},\ \bibinfo {pages} {214403} (\bibinfo {year} {2021})}\BibitemShut {NoStop}%
\bibitem [{\citenamefont {Song}\ \emph {et~al.}(2023)\citenamefont {Song}, \citenamefont {Seifert}, \citenamefont {Luo},\ and\ \citenamefont {Balents}}]{sszb23}%
  \BibitemOpen
  \bibfield  {author} {\bibinfo {author} {\bibfnamefont {Z.}~\bibnamefont {Song}}, \bibinfo {author} {\bibfnamefont {U.~F.~P.}\ \bibnamefont {Seifert}}, \bibinfo {author} {\bibfnamefont {Z.-X.}\ \bibnamefont {Luo}},\ and\ \bibinfo {author} {\bibfnamefont {L.}~\bibnamefont {Balents}},\ }\href {https://doi.org/10.1103/PhysRevB.108.155109} {\bibfield  {journal} {\bibinfo  {journal} {Phys. Rev. B}\ }\textbf {\bibinfo {volume} {108}},\ \bibinfo {pages} {155109} (\bibinfo {year} {2023})}\BibitemShut {NoStop}%
\bibitem [{\citenamefont {Rossi}\ \emph {et~al.}(2023)\citenamefont {Rossi}, \citenamefont {Motruk}, \citenamefont {Rademaker},\ and\ \citenamefont {Abanin}}]{rossi23}%
  \BibitemOpen
  \bibfield  {author} {\bibinfo {author} {\bibfnamefont {D.}~\bibnamefont {Rossi}}, \bibinfo {author} {\bibfnamefont {J.}~\bibnamefont {Motruk}}, \bibinfo {author} {\bibfnamefont {L.}~\bibnamefont {Rademaker}},\ and\ \bibinfo {author} {\bibfnamefont {D.~A.}\ \bibnamefont {Abanin}},\ }\href {https://doi.org/10.1103/PhysRevB.108.144406} {\bibfield  {journal} {\bibinfo  {journal} {Phys. Rev. B}\ }\textbf {\bibinfo {volume} {108}},\ \bibinfo {pages} {144406} (\bibinfo {year} {2023})}\BibitemShut {NoStop}%
\bibitem [{\citenamefont {Motruk}\ \emph {et~al.}(2023)\citenamefont {Motruk}, \citenamefont {Rossi}, \citenamefont {Abanin},\ and\ \citenamefont {Rademaker}}]{motruk23}%
  \BibitemOpen
  \bibfield  {author} {\bibinfo {author} {\bibfnamefont {J.}~\bibnamefont {Motruk}}, \bibinfo {author} {\bibfnamefont {D.}~\bibnamefont {Rossi}}, \bibinfo {author} {\bibfnamefont {D.~A.}\ \bibnamefont {Abanin}},\ and\ \bibinfo {author} {\bibfnamefont {L.}~\bibnamefont {Rademaker}},\ }\href {https://doi.org/10.1103/PhysRevResearch.5.L022049} {\bibfield  {journal} {\bibinfo  {journal} {Phys. Rev. Res.}\ }\textbf {\bibinfo {volume} {5}},\ \bibinfo {pages} {L022049} (\bibinfo {year} {2023})}\BibitemShut {NoStop}%
\bibitem [{\citenamefont {Zhou}\ \emph {et~al.}(2024)\citenamefont {Zhou}, \citenamefont {Sheng},\ and\ \citenamefont {Kim}}]{zhou2024quantum}%
  \BibitemOpen
  \bibfield  {author} {\bibinfo {author} {\bibfnamefont {Y.}~\bibnamefont {Zhou}}, \bibinfo {author} {\bibfnamefont {D.~N.}\ \bibnamefont {Sheng}},\ and\ \bibinfo {author} {\bibfnamefont {E.-A.}\ \bibnamefont {Kim}},\ }\href {https://doi.org/10.1103/PhysRevLett.133.156501} {\bibfield  {journal} {\bibinfo  {journal} {Phys. Rev. Lett.}\ }\textbf {\bibinfo {volume} {133}},\ \bibinfo {pages} {156501} (\bibinfo {year} {2024})}\BibitemShut {NoStop}%
\bibitem [{\citenamefont {Biborski}\ and\ \citenamefont {Zegrodnik}(2025)}]{biborski24}%
  \BibitemOpen
  \bibfield  {author} {\bibinfo {author} {\bibfnamefont {A.}~\bibnamefont {Biborski}}\ and\ \bibinfo {author} {\bibfnamefont {M.}~\bibnamefont {Zegrodnik}},\ }\href {https://doi.org/10.1103/PhysRevB.111.075116} {\bibfield  {journal} {\bibinfo  {journal} {Phys. Rev. B}\ }\textbf {\bibinfo {volume} {111}},\ \bibinfo {pages} {075116} (\bibinfo {year} {2025})}\BibitemShut {NoStop}%
\bibitem [{\citenamefont {Mak}\ and\ \citenamefont {Shan}(2022)}]{mak22}%
  \BibitemOpen
  \bibfield  {author} {\bibinfo {author} {\bibfnamefont {K.~F.}\ \bibnamefont {Mak}}\ and\ \bibinfo {author} {\bibfnamefont {J.}~\bibnamefont {Shan}},\ }\href {https://doi.org/10.1038/s41565-022-01165-6} {\bibfield  {journal} {\bibinfo  {journal} {Nature Nanotechnology}\ }\textbf {\bibinfo {volume} {17}},\ \bibinfo {pages} {686} (\bibinfo {year} {2022})}\BibitemShut {NoStop}%
\bibitem [{\citenamefont {Yan}\ \emph {et~al.}(2011{\natexlab{a}})\citenamefont {Yan}, \citenamefont {Huse},\ and\ \citenamefont {White}}]{Yan2011}%
  \BibitemOpen
  \bibfield  {author} {\bibinfo {author} {\bibfnamefont {S.}~\bibnamefont {Yan}}, \bibinfo {author} {\bibfnamefont {D.}~\bibnamefont {Huse}},\ and\ \bibinfo {author} {\bibfnamefont {S.}~\bibnamefont {White}},\ }\href@noop {} {\bibfield  {journal} {\bibinfo  {journal} {Science}\ }\textbf {\bibinfo {volume} {332}},\ \bibinfo {pages} {1173} (\bibinfo {year} {2011}{\natexlab{a}})}\BibitemShut {NoStop}%
\bibitem [{\citenamefont {Xu}\ \emph {et~al.}(2020)\citenamefont {Xu}, \citenamefont {Liu}, \citenamefont {Rhodes}, \citenamefont {Watanabe}, \citenamefont {Taniguchi}, \citenamefont {Hone}, \citenamefont {Elser}, \citenamefont {Mak},\ and\ \citenamefont {Shan}}]{xu20}%
  \BibitemOpen
  \bibfield  {author} {\bibinfo {author} {\bibfnamefont {Y.}~\bibnamefont {Xu}}, \bibinfo {author} {\bibfnamefont {S.}~\bibnamefont {Liu}}, \bibinfo {author} {\bibfnamefont {D.~A.}\ \bibnamefont {Rhodes}}, \bibinfo {author} {\bibfnamefont {K.}~\bibnamefont {Watanabe}}, \bibinfo {author} {\bibfnamefont {T.}~\bibnamefont {Taniguchi}}, \bibinfo {author} {\bibfnamefont {J.}~\bibnamefont {Hone}}, \bibinfo {author} {\bibfnamefont {V.}~\bibnamefont {Elser}}, \bibinfo {author} {\bibfnamefont {K.~F.}\ \bibnamefont {Mak}},\ and\ \bibinfo {author} {\bibfnamefont {J.}~\bibnamefont {Shan}},\ }\href {https://doi.org/10.1038/s41586-020-2868-6} {\bibfield  {journal} {\bibinfo  {journal} {Nature}\ }\textbf {\bibinfo {volume} {587}},\ \bibinfo {pages} {214} (\bibinfo {year} {2020})}\BibitemShut {NoStop}%
\bibitem [{\citenamefont {Li}\ \emph {et~al.}(2021{\natexlab{c}})\citenamefont {Li}, \citenamefont {Zhu}, \citenamefont {Tang}, \citenamefont {Watanabe}, \citenamefont {Taniguchi}, \citenamefont {Elser}, \citenamefont {Shan},\ and\ \citenamefont {Mak}}]{tingxin21}%
  \BibitemOpen
  \bibfield  {author} {\bibinfo {author} {\bibfnamefont {T.}~\bibnamefont {Li}}, \bibinfo {author} {\bibfnamefont {J.}~\bibnamefont {Zhu}}, \bibinfo {author} {\bibfnamefont {Y.}~\bibnamefont {Tang}}, \bibinfo {author} {\bibfnamefont {K.}~\bibnamefont {Watanabe}}, \bibinfo {author} {\bibfnamefont {T.}~\bibnamefont {Taniguchi}}, \bibinfo {author} {\bibfnamefont {V.}~\bibnamefont {Elser}}, \bibinfo {author} {\bibfnamefont {J.}~\bibnamefont {Shan}},\ and\ \bibinfo {author} {\bibfnamefont {K.~F.}\ \bibnamefont {Mak}},\ }\href {https://doi.org/10.1038/s41565-021-00955-8} {\bibfield  {journal} {\bibinfo  {journal} {Nature Nanotechnology}\ }\textbf {\bibinfo {volume} {16}},\ \bibinfo {pages} {1068} (\bibinfo {year} {2021}{\natexlab{c}})}\BibitemShut {NoStop}%
\bibitem [{\citenamefont {Rastelli}\ \emph {et~al.}(1979)\citenamefont {Rastelli}, \citenamefont {Tassi},\ and\ \citenamefont {Reatto}}]{rastelli79}%
  \BibitemOpen
  \bibfield  {author} {\bibinfo {author} {\bibfnamefont {E.}~\bibnamefont {Rastelli}}, \bibinfo {author} {\bibfnamefont {A.}~\bibnamefont {Tassi}},\ and\ \bibinfo {author} {\bibfnamefont {L.}~\bibnamefont {Reatto}},\ }\href {https://doi.org/https://doi.org/10.1016/0378-4363(79)90002-0} {\bibfield  {journal} {\bibinfo  {journal} {Physica B+C}\ }\textbf {\bibinfo {volume} {97}},\ \bibinfo {pages} {1} (\bibinfo {year} {1979})}\BibitemShut {NoStop}%
\bibitem [{\citenamefont {Fouet}\ \emph {et~al.}(2001)\citenamefont {Fouet}, \citenamefont {Sindzingre},\ and\ \citenamefont {Lhuillier}}]{fouet01}%
  \BibitemOpen
  \bibfield  {author} {\bibinfo {author} {\bibfnamefont {J.~B.}\ \bibnamefont {Fouet}}, \bibinfo {author} {\bibfnamefont {P.}~\bibnamefont {Sindzingre}},\ and\ \bibinfo {author} {\bibfnamefont {C.}~\bibnamefont {Lhuillier}},\ }\href {https://doi.org/10.1007/s100510170273} {\bibfield  {journal} {\bibinfo  {journal} {The European Physical Journal B - Condensed Matter and Complex Systems}\ }\textbf {\bibinfo {volume} {20}},\ \bibinfo {pages} {241} (\bibinfo {year} {2001})}\BibitemShut {NoStop}%
\bibitem [{\citenamefont {Davoudi}\ \emph {et~al.}(2008)\citenamefont {Davoudi}, \citenamefont {Hassan},\ and\ \citenamefont {Tremblay}}]{davoudi08}%
  \BibitemOpen
  \bibfield  {author} {\bibinfo {author} {\bibfnamefont {B.}~\bibnamefont {Davoudi}}, \bibinfo {author} {\bibfnamefont {S.~R.}\ \bibnamefont {Hassan}},\ and\ \bibinfo {author} {\bibfnamefont {A.-M.~S.}\ \bibnamefont {Tremblay}},\ }\href {https://doi.org/10.1103/PhysRevB.77.214408} {\bibfield  {journal} {\bibinfo  {journal} {Phys. Rev. B}\ }\textbf {\bibinfo {volume} {77}},\ \bibinfo {pages} {214408} (\bibinfo {year} {2008})}\BibitemShut {NoStop}%
\bibitem [{\citenamefont {Lieb}\ \emph {et~al.}(1961)\citenamefont {Lieb}, \citenamefont {Schultz},\ and\ \citenamefont {Mattis}}]{lieb61}%
  \BibitemOpen
  \bibfield  {author} {\bibinfo {author} {\bibfnamefont {E.}~\bibnamefont {Lieb}}, \bibinfo {author} {\bibfnamefont {T.}~\bibnamefont {Schultz}},\ and\ \bibinfo {author} {\bibfnamefont {D.}~\bibnamefont {Mattis}},\ }\href {https://doi.org/https://doi.org/10.1016/0003-4916(61)90115-4} {\bibfield  {journal} {\bibinfo  {journal} {Annals of Physics}\ }\textbf {\bibinfo {volume} {16}},\ \bibinfo {pages} {407} (\bibinfo {year} {1961})}\BibitemShut {NoStop}%
\bibitem [{\citenamefont {Oshikawa}(2000)}]{oshikawa00}%
  \BibitemOpen
  \bibfield  {author} {\bibinfo {author} {\bibfnamefont {M.}~\bibnamefont {Oshikawa}},\ }\href {https://doi.org/10.1103/PhysRevLett.84.1535} {\bibfield  {journal} {\bibinfo  {journal} {Phys. Rev. Lett.}\ }\textbf {\bibinfo {volume} {84}},\ \bibinfo {pages} {1535} (\bibinfo {year} {2000})}\BibitemShut {NoStop}%
\bibitem [{\citenamefont {Saadatmand}\ \emph {et~al.}(2015)\citenamefont {Saadatmand}, \citenamefont {Powell},\ and\ \citenamefont {McCulloch}}]{saad15}%
  \BibitemOpen
  \bibfield  {author} {\bibinfo {author} {\bibfnamefont {S.~N.}\ \bibnamefont {Saadatmand}}, \bibinfo {author} {\bibfnamefont {B.~J.}\ \bibnamefont {Powell}},\ and\ \bibinfo {author} {\bibfnamefont {I.~P.}\ \bibnamefont {McCulloch}},\ }\href {https://doi.org/10.1103/PhysRevB.91.245119} {\bibfield  {journal} {\bibinfo  {journal} {Phys. Rev. B}\ }\textbf {\bibinfo {volume} {91}},\ \bibinfo {pages} {245119} (\bibinfo {year} {2015})}\BibitemShut {NoStop}%
\bibitem [{\citenamefont {Zhu}\ and\ \citenamefont {White}(2015)}]{zhu15}%
  \BibitemOpen
  \bibfield  {author} {\bibinfo {author} {\bibfnamefont {Z.}~\bibnamefont {Zhu}}\ and\ \bibinfo {author} {\bibfnamefont {S.~R.}\ \bibnamefont {White}},\ }\href {https://doi.org/10.1103/PhysRevB.92.041105} {\bibfield  {journal} {\bibinfo  {journal} {Phys. Rev. B}\ }\textbf {\bibinfo {volume} {92}},\ \bibinfo {pages} {041105} (\bibinfo {year} {2015})}\BibitemShut {NoStop}%
\bibitem [{\citenamefont {Chen}\ \emph {et~al.}(2013)\citenamefont {Chen}, \citenamefont {Ju}, \citenamefont {Jiang}, \citenamefont {Starykh},\ and\ \citenamefont {Balents}}]{chen13}%
  \BibitemOpen
  \bibfield  {author} {\bibinfo {author} {\bibfnamefont {R.}~\bibnamefont {Chen}}, \bibinfo {author} {\bibfnamefont {H.}~\bibnamefont {Ju}}, \bibinfo {author} {\bibfnamefont {H.-C.}\ \bibnamefont {Jiang}}, \bibinfo {author} {\bibfnamefont {O.~A.}\ \bibnamefont {Starykh}},\ and\ \bibinfo {author} {\bibfnamefont {L.}~\bibnamefont {Balents}},\ }\href {https://doi.org/10.1103/PhysRevB.87.165123} {\bibfield  {journal} {\bibinfo  {journal} {Phys. Rev. B}\ }\textbf {\bibinfo {volume} {87}},\ \bibinfo {pages} {165123} (\bibinfo {year} {2013})}\BibitemShut {NoStop}%
\bibitem [{\citenamefont {Peng}\ \emph {et~al.}(2021)\citenamefont {Peng}, \citenamefont {Jiang}, \citenamefont {Wang},\ and\ \citenamefont {Jiang}}]{peng21}%
  \BibitemOpen
  \bibfield  {author} {\bibinfo {author} {\bibfnamefont {C.}~\bibnamefont {Peng}}, \bibinfo {author} {\bibfnamefont {Y.-F.}\ \bibnamefont {Jiang}}, \bibinfo {author} {\bibfnamefont {Y.}~\bibnamefont {Wang}},\ and\ \bibinfo {author} {\bibfnamefont {H.-C.}\ \bibnamefont {Jiang}},\ }\href {https://doi.org/10.1088/1367-2630/ac3a83} {\bibfield  {journal} {\bibinfo  {journal} {New Journal of Physics}\ }\textbf {\bibinfo {volume} {23}},\ \bibinfo {pages} {123004} (\bibinfo {year} {2021})}\BibitemShut {NoStop}%
\bibitem [{\citenamefont {Szasz}\ \emph {et~al.}(2020)\citenamefont {Szasz}, \citenamefont {Motruk}, \citenamefont {Zaletel},\ and\ \citenamefont {Moore}}]{szasz20}%
  \BibitemOpen
  \bibfield  {author} {\bibinfo {author} {\bibfnamefont {A.}~\bibnamefont {Szasz}}, \bibinfo {author} {\bibfnamefont {J.}~\bibnamefont {Motruk}}, \bibinfo {author} {\bibfnamefont {M.~P.}\ \bibnamefont {Zaletel}},\ and\ \bibinfo {author} {\bibfnamefont {J.~E.}\ \bibnamefont {Moore}},\ }\href {https://doi.org/10.1103/PhysRevX.10.021042} {\bibfield  {journal} {\bibinfo  {journal} {Phys. Rev. X}\ }\textbf {\bibinfo {volume} {10}},\ \bibinfo {pages} {021042} (\bibinfo {year} {2020})}\BibitemShut {NoStop}%
\bibitem [{\citenamefont {Wietek}\ \emph {et~al.}(2021)\citenamefont {Wietek}, \citenamefont {Rossi}, \citenamefont {\ifmmode~\check{S}\else \v{S}\fi{}imkovic}, \citenamefont {Klett}, \citenamefont {Hansmann}, \citenamefont {Ferrero}, \citenamefont {Stoudenmire}, \citenamefont {Sch\"afer},\ and\ \citenamefont {Georges}}]{wietek21}%
  \BibitemOpen
  \bibfield  {author} {\bibinfo {author} {\bibfnamefont {A.}~\bibnamefont {Wietek}}, \bibinfo {author} {\bibfnamefont {R.}~\bibnamefont {Rossi}}, \bibinfo {author} {\bibfnamefont {F.}~\bibnamefont {\ifmmode~\check{S}\else \v{S}\fi{}imkovic}}, \bibinfo {author} {\bibfnamefont {M.}~\bibnamefont {Klett}}, \bibinfo {author} {\bibfnamefont {P.}~\bibnamefont {Hansmann}}, \bibinfo {author} {\bibfnamefont {M.}~\bibnamefont {Ferrero}}, \bibinfo {author} {\bibfnamefont {E.~M.}\ \bibnamefont {Stoudenmire}}, \bibinfo {author} {\bibfnamefont {T.}~\bibnamefont {Sch\"afer}},\ and\ \bibinfo {author} {\bibfnamefont {A.}~\bibnamefont {Georges}},\ }\href {https://doi.org/10.1103/PhysRevX.11.041013} {\bibfield  {journal} {\bibinfo  {journal} {Phys. Rev. X}\ }\textbf {\bibinfo {volume} {11}},\ \bibinfo {pages} {041013} (\bibinfo {year} {2021})}\BibitemShut {NoStop}%
\bibitem [{\citenamefont {Messio}\ \emph {et~al.}(2010)\citenamefont {Messio}, \citenamefont {C\'epas},\ and\ \citenamefont {Lhuillier}}]{messio10}%
  \BibitemOpen
  \bibfield  {author} {\bibinfo {author} {\bibfnamefont {L.}~\bibnamefont {Messio}}, \bibinfo {author} {\bibfnamefont {O.}~\bibnamefont {C\'epas}},\ and\ \bibinfo {author} {\bibfnamefont {C.}~\bibnamefont {Lhuillier}},\ }\href {https://doi.org/10.1103/PhysRevB.81.064428} {\bibfield  {journal} {\bibinfo  {journal} {Phys. Rev. B}\ }\textbf {\bibinfo {volume} {81}},\ \bibinfo {pages} {064428} (\bibinfo {year} {2010})}\BibitemShut {NoStop}%
\bibitem [{\citenamefont {Iqbal}\ \emph {et~al.}(2013)\citenamefont {Iqbal}, \citenamefont {Becca}, \citenamefont {Sorella},\ and\ \citenamefont {Poilblanc}}]{iqbal13}%
  \BibitemOpen
  \bibfield  {author} {\bibinfo {author} {\bibfnamefont {Y.}~\bibnamefont {Iqbal}}, \bibinfo {author} {\bibfnamefont {F.}~\bibnamefont {Becca}}, \bibinfo {author} {\bibfnamefont {S.}~\bibnamefont {Sorella}},\ and\ \bibinfo {author} {\bibfnamefont {D.}~\bibnamefont {Poilblanc}},\ }\href {https://doi.org/10.1103/PhysRevB.87.060405} {\bibfield  {journal} {\bibinfo  {journal} {Phys. Rev. B}\ }\textbf {\bibinfo {volume} {87}},\ \bibinfo {pages} {060405} (\bibinfo {year} {2013})}\BibitemShut {NoStop}%
\bibitem [{\citenamefont {Punk}\ \emph {et~al.}(2014)\citenamefont {Punk}, \citenamefont {Chowdhury},\ and\ \citenamefont {Sachdev}}]{punk14}%
  \BibitemOpen
  \bibfield  {author} {\bibinfo {author} {\bibfnamefont {M.}~\bibnamefont {Punk}}, \bibinfo {author} {\bibfnamefont {D.}~\bibnamefont {Chowdhury}},\ and\ \bibinfo {author} {\bibfnamefont {S.}~\bibnamefont {Sachdev}},\ }\href {https://doi.org/10.1038/nphys2887} {\bibfield  {journal} {\bibinfo  {journal} {Nature Physics}\ }\textbf {\bibinfo {volume} {10}},\ \bibinfo {pages} {289} (\bibinfo {year} {2014})}\BibitemShut {NoStop}%
\bibitem [{\citenamefont {Fu}\ \emph {et~al.}(2015)\citenamefont {Fu}, \citenamefont {Imai}, \citenamefont {Han},\ and\ \citenamefont {Lee}}]{Fu2015}%
  \BibitemOpen
  \bibfield  {author} {\bibinfo {author} {\bibfnamefont {M.}~\bibnamefont {Fu}}, \bibinfo {author} {\bibfnamefont {T.}~\bibnamefont {Imai}}, \bibinfo {author} {\bibfnamefont {T.-H.}\ \bibnamefont {Han}},\ and\ \bibinfo {author} {\bibfnamefont {Y.~S.}\ \bibnamefont {Lee}},\ }\href {https://doi.org/10.1126/science.aab2120} {\bibfield  {journal} {\bibinfo  {journal} {Science}\ }\textbf {\bibinfo {volume} {350}},\ \bibinfo {pages} {655} (\bibinfo {year} {2015})}\BibitemShut {NoStop}%
\bibitem [{\citenamefont {Kolley}\ \emph {et~al.}(2015)\citenamefont {Kolley}, \citenamefont {Depenbrock}, \citenamefont {McCulloch}, \citenamefont {Schollw\"ock},\ and\ \citenamefont {Alba}}]{kolley15}%
  \BibitemOpen
  \bibfield  {author} {\bibinfo {author} {\bibfnamefont {F.}~\bibnamefont {Kolley}}, \bibinfo {author} {\bibfnamefont {S.}~\bibnamefont {Depenbrock}}, \bibinfo {author} {\bibfnamefont {I.~P.}\ \bibnamefont {McCulloch}}, \bibinfo {author} {\bibfnamefont {U.}~\bibnamefont {Schollw\"ock}},\ and\ \bibinfo {author} {\bibfnamefont {V.}~\bibnamefont {Alba}},\ }\href {https://doi.org/10.1103/PhysRevB.91.104418} {\bibfield  {journal} {\bibinfo  {journal} {Phys. Rev. B}\ }\textbf {\bibinfo {volume} {91}},\ \bibinfo {pages} {104418} (\bibinfo {year} {2015})}\BibitemShut {NoStop}%
\bibitem [{\citenamefont {Halimeh}\ and\ \citenamefont {Punk}(2016)}]{halimeh16}%
  \BibitemOpen
  \bibfield  {author} {\bibinfo {author} {\bibfnamefont {J.~C.}\ \bibnamefont {Halimeh}}\ and\ \bibinfo {author} {\bibfnamefont {M.}~\bibnamefont {Punk}},\ }\href {https://doi.org/10.1103/PhysRevB.94.104413} {\bibfield  {journal} {\bibinfo  {journal} {Phys. Rev. B}\ }\textbf {\bibinfo {volume} {94}},\ \bibinfo {pages} {104413} (\bibinfo {year} {2016})}\BibitemShut {NoStop}%
\bibitem [{\citenamefont {Messio}\ \emph {et~al.}(2017)\citenamefont {Messio}, \citenamefont {Bieri}, \citenamefont {Lhuillier},\ and\ \citenamefont {Bernu}}]{messio17}%
  \BibitemOpen
  \bibfield  {author} {\bibinfo {author} {\bibfnamefont {L.}~\bibnamefont {Messio}}, \bibinfo {author} {\bibfnamefont {S.}~\bibnamefont {Bieri}}, \bibinfo {author} {\bibfnamefont {C.}~\bibnamefont {Lhuillier}},\ and\ \bibinfo {author} {\bibfnamefont {B.}~\bibnamefont {Bernu}},\ }\href {https://doi.org/10.1103/PhysRevLett.118.267201} {\bibfield  {journal} {\bibinfo  {journal} {Phys. Rev. Lett.}\ }\textbf {\bibinfo {volume} {118}},\ \bibinfo {pages} {267201} (\bibinfo {year} {2017})}\BibitemShut {NoStop}%
\bibitem [{\citenamefont {Chern}\ and\ \citenamefont {Kim}(2021)}]{chern21}%
  \BibitemOpen
  \bibfield  {author} {\bibinfo {author} {\bibfnamefont {L.~E.}\ \bibnamefont {Chern}}\ and\ \bibinfo {author} {\bibfnamefont {Y.~B.}\ \bibnamefont {Kim}},\ }\href {https://doi.org/10.1103/PhysRevB.104.094413} {\bibfield  {journal} {\bibinfo  {journal} {Phys. Rev. B}\ }\textbf {\bibinfo {volume} {104}},\ \bibinfo {pages} {094413} (\bibinfo {year} {2021})}\BibitemShut {NoStop}%
\bibitem [{\citenamefont {He}\ \emph {et~al.}(2017)\citenamefont {He}, \citenamefont {Zaletel}, \citenamefont {Oshikawa},\ and\ \citenamefont {Pollmann}}]{he17}%
  \BibitemOpen
  \bibfield  {author} {\bibinfo {author} {\bibfnamefont {Y.-C.}\ \bibnamefont {He}}, \bibinfo {author} {\bibfnamefont {M.~P.}\ \bibnamefont {Zaletel}}, \bibinfo {author} {\bibfnamefont {M.}~\bibnamefont {Oshikawa}},\ and\ \bibinfo {author} {\bibfnamefont {F.}~\bibnamefont {Pollmann}},\ }\href {https://doi.org/10.1103/PhysRevX.7.031020} {\bibfield  {journal} {\bibinfo  {journal} {Phys. Rev. X}\ }\textbf {\bibinfo {volume} {7}},\ \bibinfo {pages} {031020} (\bibinfo {year} {2017})}\BibitemShut {NoStop}%
\bibitem [{\citenamefont {Hu}\ \emph {et~al.}(2019)\citenamefont {Hu}, \citenamefont {Zhu}, \citenamefont {Eggert},\ and\ \citenamefont {He}}]{hu19}%
  \BibitemOpen
  \bibfield  {author} {\bibinfo {author} {\bibfnamefont {S.}~\bibnamefont {Hu}}, \bibinfo {author} {\bibfnamefont {W.}~\bibnamefont {Zhu}}, \bibinfo {author} {\bibfnamefont {S.}~\bibnamefont {Eggert}},\ and\ \bibinfo {author} {\bibfnamefont {Y.-C.}\ \bibnamefont {He}},\ }\href {https://doi.org/10.1103/PhysRevLett.123.207203} {\bibfield  {journal} {\bibinfo  {journal} {Phys. Rev. Lett.}\ }\textbf {\bibinfo {volume} {123}},\ \bibinfo {pages} {207203} (\bibinfo {year} {2019})}\BibitemShut {NoStop}%
\bibitem [{\citenamefont {Hermele}\ \emph {et~al.}(2008)\citenamefont {Hermele}, \citenamefont {Ran}, \citenamefont {Lee},\ and\ \citenamefont {Wen}}]{hermele08}%
  \BibitemOpen
  \bibfield  {author} {\bibinfo {author} {\bibfnamefont {M.}~\bibnamefont {Hermele}}, \bibinfo {author} {\bibfnamefont {Y.}~\bibnamefont {Ran}}, \bibinfo {author} {\bibfnamefont {P.~A.}\ \bibnamefont {Lee}},\ and\ \bibinfo {author} {\bibfnamefont {X.-G.}\ \bibnamefont {Wen}},\ }\href {https://doi.org/10.1103/PhysRevB.77.224413} {\bibfield  {journal} {\bibinfo  {journal} {Phys. Rev. B}\ }\textbf {\bibinfo {volume} {77}},\ \bibinfo {pages} {224413} (\bibinfo {year} {2008})}\BibitemShut {NoStop}%
\bibitem [{\citenamefont {Iqbal}\ \emph {et~al.}(2014)\citenamefont {Iqbal}, \citenamefont {Poilblanc},\ and\ \citenamefont {Becca}}]{iqbal14}%
  \BibitemOpen
  \bibfield  {author} {\bibinfo {author} {\bibfnamefont {Y.}~\bibnamefont {Iqbal}}, \bibinfo {author} {\bibfnamefont {D.}~\bibnamefont {Poilblanc}},\ and\ \bibinfo {author} {\bibfnamefont {F.}~\bibnamefont {Becca}},\ }\href {https://doi.org/10.1103/PhysRevB.89.020407} {\bibfield  {journal} {\bibinfo  {journal} {Phys. Rev. B}\ }\textbf {\bibinfo {volume} {89}},\ \bibinfo {pages} {020407} (\bibinfo {year} {2014})}\BibitemShut {NoStop}%
\bibitem [{\citenamefont {Iqbal}\ \emph {et~al.}(2015)\citenamefont {Iqbal}, \citenamefont {Poilblanc},\ and\ \citenamefont {Becca}}]{iqbal15}%
  \BibitemOpen
  \bibfield  {author} {\bibinfo {author} {\bibfnamefont {Y.}~\bibnamefont {Iqbal}}, \bibinfo {author} {\bibfnamefont {D.}~\bibnamefont {Poilblanc}},\ and\ \bibinfo {author} {\bibfnamefont {F.}~\bibnamefont {Becca}},\ }\href {https://doi.org/10.1103/PhysRevB.91.020402} {\bibfield  {journal} {\bibinfo  {journal} {Phys. Rev. B}\ }\textbf {\bibinfo {volume} {91}},\ \bibinfo {pages} {020402} (\bibinfo {year} {2015})}\BibitemShut {NoStop}%
\bibitem [{\citenamefont {Zhu}\ \emph {et~al.}(2018)\citenamefont {Zhu}, \citenamefont {Chen}, \citenamefont {He},\ and\ \citenamefont {Witczak-Krempa}}]{zhu18}%
  \BibitemOpen
  \bibfield  {author} {\bibinfo {author} {\bibfnamefont {W.}~\bibnamefont {Zhu}}, \bibinfo {author} {\bibfnamefont {X.}~\bibnamefont {Chen}}, \bibinfo {author} {\bibfnamefont {Y.-C.}\ \bibnamefont {He}},\ and\ \bibinfo {author} {\bibfnamefont {W.}~\bibnamefont {Witczak-Krempa}},\ }\href {https://doi.org/10.1126/sciadv.aat5535} {\bibfield  {journal} {\bibinfo  {journal} {Science Advances}\ }\textbf {\bibinfo {volume} {4}},\ \bibinfo {pages} {eaat5535} (\bibinfo {year} {2018})}\BibitemShut {NoStop}%
\bibitem [{\citenamefont {Han}\ \emph {et~al.}(2012)\citenamefont {Han}, \citenamefont {Helton}, \citenamefont {Chu}, \citenamefont {Nocera}, \citenamefont {Rodriguez-Rivera}, \citenamefont {Broholm},\ and\ \citenamefont {Lee}}]{han12}%
  \BibitemOpen
  \bibfield  {author} {\bibinfo {author} {\bibfnamefont {T.-H.}\ \bibnamefont {Han}}, \bibinfo {author} {\bibfnamefont {J.~S.}\ \bibnamefont {Helton}}, \bibinfo {author} {\bibfnamefont {S.}~\bibnamefont {Chu}}, \bibinfo {author} {\bibfnamefont {D.~G.}\ \bibnamefont {Nocera}}, \bibinfo {author} {\bibfnamefont {J.~A.}\ \bibnamefont {Rodriguez-Rivera}}, \bibinfo {author} {\bibfnamefont {C.}~\bibnamefont {Broholm}},\ and\ \bibinfo {author} {\bibfnamefont {Y.~S.}\ \bibnamefont {Lee}},\ }\href {https://doi.org/10.1038/nature11659} {\bibfield  {journal} {\bibinfo  {journal} {Nature}\ }\textbf {\bibinfo {volume} {492}},\ \bibinfo {pages} {406} (\bibinfo {year} {2012})}\BibitemShut {NoStop}%
\bibitem [{\citenamefont {F\aa{}k}\ \emph {et~al.}(2012)\citenamefont {F\aa{}k}, \citenamefont {Kermarrec}, \citenamefont {Messio}, \citenamefont {Bernu}, \citenamefont {Lhuillier}, \citenamefont {Bert}, \citenamefont {Mendels}, \citenamefont {Koteswararao}, \citenamefont {Bouquet}, \citenamefont {Ollivier}, \citenamefont {Hillier}, \citenamefont {Amato}, \citenamefont {Colman},\ and\ \citenamefont {Wills}}]{fak12}%
  \BibitemOpen
  \bibfield  {author} {\bibinfo {author} {\bibfnamefont {B.}~\bibnamefont {F\aa{}k}}, \bibinfo {author} {\bibfnamefont {E.}~\bibnamefont {Kermarrec}}, \bibinfo {author} {\bibfnamefont {L.}~\bibnamefont {Messio}}, \bibinfo {author} {\bibfnamefont {B.}~\bibnamefont {Bernu}}, \bibinfo {author} {\bibfnamefont {C.}~\bibnamefont {Lhuillier}}, \bibinfo {author} {\bibfnamefont {F.}~\bibnamefont {Bert}}, \bibinfo {author} {\bibfnamefont {P.}~\bibnamefont {Mendels}}, \bibinfo {author} {\bibfnamefont {B.}~\bibnamefont {Koteswararao}}, \bibinfo {author} {\bibfnamefont {F.}~\bibnamefont {Bouquet}}, \bibinfo {author} {\bibfnamefont {J.}~\bibnamefont {Ollivier}}, \bibinfo {author} {\bibfnamefont {A.~D.}\ \bibnamefont {Hillier}}, \bibinfo {author} {\bibfnamefont {A.}~\bibnamefont {Amato}}, \bibinfo {author} {\bibfnamefont {R.~H.}\ \bibnamefont {Colman}},\ and\ \bibinfo {author} {\bibfnamefont {A.~S.}\ \bibnamefont {Wills}},\ }\href {https://doi.org/10.1103/PhysRevLett.109.037208} {\bibfield  {journal} {\bibinfo  {journal}
  {Phys. Rev. Lett.}\ }\textbf {\bibinfo {volume} {109}},\ \bibinfo {pages} {037208} (\bibinfo {year} {2012})}\BibitemShut {NoStop}%
\bibitem [{\citenamefont {Jiang}\ \emph {et~al.}(2008)\citenamefont {Jiang}, \citenamefont {Weng},\ and\ \citenamefont {Sheng}}]{jiang08}%
  \BibitemOpen
  \bibfield  {author} {\bibinfo {author} {\bibfnamefont {H.~C.}\ \bibnamefont {Jiang}}, \bibinfo {author} {\bibfnamefont {Z.~Y.}\ \bibnamefont {Weng}},\ and\ \bibinfo {author} {\bibfnamefont {D.~N.}\ \bibnamefont {Sheng}},\ }\href {https://doi.org/10.1103/PhysRevLett.101.117203} {\bibfield  {journal} {\bibinfo  {journal} {Phys. Rev. Lett.}\ }\textbf {\bibinfo {volume} {101}},\ \bibinfo {pages} {117203} (\bibinfo {year} {2008})}\BibitemShut {NoStop}%
\bibitem [{\citenamefont {Yan}\ \emph {et~al.}(2011{\natexlab{b}})\citenamefont {Yan}, \citenamefont {Huse},\ and\ \citenamefont {White}}]{yan11}%
  \BibitemOpen
  \bibfield  {author} {\bibinfo {author} {\bibfnamefont {S.}~\bibnamefont {Yan}}, \bibinfo {author} {\bibfnamefont {D.~A.}\ \bibnamefont {Huse}},\ and\ \bibinfo {author} {\bibfnamefont {S.~R.}\ \bibnamefont {White}},\ }\href {https://doi.org/10.1126/science.1201080} {\bibfield  {journal} {\bibinfo  {journal} {Science}\ }\textbf {\bibinfo {volume} {332}},\ \bibinfo {pages} {1173} (\bibinfo {year} {2011}{\natexlab{b}})}\BibitemShut {NoStop}%
\bibitem [{\citenamefont {Jiang}\ \emph {et~al.}(2012)\citenamefont {Jiang}, \citenamefont {Wang},\ and\ \citenamefont {Balents}}]{jiang12}%
  \BibitemOpen
  \bibfield  {author} {\bibinfo {author} {\bibfnamefont {H.-C.}\ \bibnamefont {Jiang}}, \bibinfo {author} {\bibfnamefont {Z.}~\bibnamefont {Wang}},\ and\ \bibinfo {author} {\bibfnamefont {L.}~\bibnamefont {Balents}},\ }\href {https://doi.org/10.1038/nphys2465} {\bibfield  {journal} {\bibinfo  {journal} {Nature Physics}\ }\textbf {\bibinfo {volume} {8}},\ \bibinfo {pages} {902} (\bibinfo {year} {2012})}\BibitemShut {NoStop}%
\bibitem [{\citenamefont {Depenbrock}\ \emph {et~al.}(2012)\citenamefont {Depenbrock}, \citenamefont {McCulloch},\ and\ \citenamefont {Schollw\"ock}}]{depenbrock12}%
  \BibitemOpen
  \bibfield  {author} {\bibinfo {author} {\bibfnamefont {S.}~\bibnamefont {Depenbrock}}, \bibinfo {author} {\bibfnamefont {I.~P.}\ \bibnamefont {McCulloch}},\ and\ \bibinfo {author} {\bibfnamefont {U.}~\bibnamefont {Schollw\"ock}},\ }\href {https://doi.org/10.1103/PhysRevLett.109.067201} {\bibfield  {journal} {\bibinfo  {journal} {Phys. Rev. Lett.}\ }\textbf {\bibinfo {volume} {109}},\ \bibinfo {pages} {067201} (\bibinfo {year} {2012})}\BibitemShut {NoStop}%
\bibitem [{\citenamefont {Messio}\ \emph {et~al.}(2012)\citenamefont {Messio}, \citenamefont {Bernu},\ and\ \citenamefont {Lhuillier}}]{messio12}%
  \BibitemOpen
  \bibfield  {author} {\bibinfo {author} {\bibfnamefont {L.}~\bibnamefont {Messio}}, \bibinfo {author} {\bibfnamefont {B.}~\bibnamefont {Bernu}},\ and\ \bibinfo {author} {\bibfnamefont {C.}~\bibnamefont {Lhuillier}},\ }\href {https://doi.org/10.1103/PhysRevLett.108.207204} {\bibfield  {journal} {\bibinfo  {journal} {Phys. Rev. Lett.}\ }\textbf {\bibinfo {volume} {108}},\ \bibinfo {pages} {207204} (\bibinfo {year} {2012})}\BibitemShut {NoStop}%
\bibitem [{\citenamefont {Gong}\ \emph {et~al.}(2014)\citenamefont {Gong}, \citenamefont {Zhu},\ and\ \citenamefont {Sheng}}]{gong14}%
  \BibitemOpen
  \bibfield  {author} {\bibinfo {author} {\bibfnamefont {S.-S.}\ \bibnamefont {Gong}}, \bibinfo {author} {\bibfnamefont {W.}~\bibnamefont {Zhu}},\ and\ \bibinfo {author} {\bibfnamefont {D.~N.}\ \bibnamefont {Sheng}},\ }\href {https://doi.org/10.1038/srep06317} {\bibfield  {journal} {\bibinfo  {journal} {Scientific Reports}\ }\textbf {\bibinfo {volume} {4}},\ \bibinfo {pages} {6317} (\bibinfo {year} {2014})}\BibitemShut {NoStop}%
\bibitem [{\citenamefont {Bauer}\ \emph {et~al.}(2014)\citenamefont {Bauer}, \citenamefont {Cincio}, \citenamefont {Keller}, \citenamefont {Dolfi}, \citenamefont {Vidal}, \citenamefont {Trebst},\ and\ \citenamefont {Ludwig}}]{bauer14}%
  \BibitemOpen
  \bibfield  {author} {\bibinfo {author} {\bibfnamefont {B.}~\bibnamefont {Bauer}}, \bibinfo {author} {\bibfnamefont {L.}~\bibnamefont {Cincio}}, \bibinfo {author} {\bibfnamefont {B.~P.}\ \bibnamefont {Keller}}, \bibinfo {author} {\bibfnamefont {M.}~\bibnamefont {Dolfi}}, \bibinfo {author} {\bibfnamefont {G.}~\bibnamefont {Vidal}}, \bibinfo {author} {\bibfnamefont {S.}~\bibnamefont {Trebst}},\ and\ \bibinfo {author} {\bibfnamefont {A.~W.~W.}\ \bibnamefont {Ludwig}},\ }\href {https://doi.org/10.1038/ncomms6137} {\bibfield  {journal} {\bibinfo  {journal} {Nature Communications}\ }\textbf {\bibinfo {volume} {5}},\ \bibinfo {pages} {5137} (\bibinfo {year} {2014})}\BibitemShut {NoStop}%
\bibitem [{\citenamefont {Wietek}\ \emph {et~al.}(2015)\citenamefont {Wietek}, \citenamefont {Sterdyniak},\ and\ \citenamefont {L\"auchli}}]{wietek15}%
  \BibitemOpen
  \bibfield  {author} {\bibinfo {author} {\bibfnamefont {A.}~\bibnamefont {Wietek}}, \bibinfo {author} {\bibfnamefont {A.}~\bibnamefont {Sterdyniak}},\ and\ \bibinfo {author} {\bibfnamefont {A.~M.}\ \bibnamefont {L\"auchli}},\ }\href {https://doi.org/10.1103/PhysRevB.92.125122} {\bibfield  {journal} {\bibinfo  {journal} {Phys. Rev. B}\ }\textbf {\bibinfo {volume} {92}},\ \bibinfo {pages} {125122} (\bibinfo {year} {2015})}\BibitemShut {NoStop}%
\bibitem [{\citenamefont {Hu}\ \emph {et~al.}(2015)\citenamefont {Hu}, \citenamefont {Zhu}, \citenamefont {Zhang}, \citenamefont {Gong}, \citenamefont {Becca},\ and\ \citenamefont {Sheng}}]{hu15}%
  \BibitemOpen
  \bibfield  {author} {\bibinfo {author} {\bibfnamefont {W.-J.}\ \bibnamefont {Hu}}, \bibinfo {author} {\bibfnamefont {W.}~\bibnamefont {Zhu}}, \bibinfo {author} {\bibfnamefont {Y.}~\bibnamefont {Zhang}}, \bibinfo {author} {\bibfnamefont {S.}~\bibnamefont {Gong}}, \bibinfo {author} {\bibfnamefont {F.}~\bibnamefont {Becca}},\ and\ \bibinfo {author} {\bibfnamefont {D.~N.}\ \bibnamefont {Sheng}},\ }\href {https://doi.org/10.1103/PhysRevB.91.041124} {\bibfield  {journal} {\bibinfo  {journal} {Phys. Rev. B}\ }\textbf {\bibinfo {volume} {91}},\ \bibinfo {pages} {041124} (\bibinfo {year} {2015})}\BibitemShut {NoStop}%
\bibitem [{\citenamefont {Seifert}\ and\ \citenamefont {Balents}(2024)}]{ufps24}%
  \BibitemOpen
  \bibfield  {author} {\bibinfo {author} {\bibfnamefont {U.~F.~P.}\ \bibnamefont {Seifert}}\ and\ \bibinfo {author} {\bibfnamefont {L.}~\bibnamefont {Balents}},\ }\href {https://doi.org/10.1103/PhysRevLett.132.046501} {\bibfield  {journal} {\bibinfo  {journal} {Phys. Rev. Lett.}\ }\textbf {\bibinfo {volume} {132}},\ \bibinfo {pages} {046501} (\bibinfo {year} {2024})}\BibitemShut {NoStop}%
\bibitem [{\citenamefont {Morera}\ \emph {et~al.}(2023)\citenamefont {Morera}, \citenamefont {Kan\'asz-Nagy}, \citenamefont {Smolenski}, \citenamefont {Ciorciaro}, \citenamefont {Imamo\ifmmode~\breve{g}\else \u{g}\fi{}lu},\ and\ \citenamefont {Demler}}]{morera23}%
  \BibitemOpen
  \bibfield  {author} {\bibinfo {author} {\bibfnamefont {I.}~\bibnamefont {Morera}}, \bibinfo {author} {\bibfnamefont {M.}~\bibnamefont {Kan\'asz-Nagy}}, \bibinfo {author} {\bibfnamefont {T.}~\bibnamefont {Smolenski}}, \bibinfo {author} {\bibfnamefont {L.}~\bibnamefont {Ciorciaro}}, \bibinfo {author} {\bibfnamefont {A.~m.~c.}\ \bibnamefont {Imamo\ifmmode~\breve{g}\else \u{g}\fi{}lu}},\ and\ \bibinfo {author} {\bibfnamefont {E.}~\bibnamefont {Demler}},\ }\href {https://doi.org/10.1103/PhysRevResearch.5.L022048} {\bibfield  {journal} {\bibinfo  {journal} {Phys. Rev. Res.}\ }\textbf {\bibinfo {volume} {5}},\ \bibinfo {pages} {L022048} (\bibinfo {year} {2023})}\BibitemShut {NoStop}%
\bibitem [{\citenamefont {Zhao}\ \emph {et~al.}(2024)\citenamefont {Zhao}, \citenamefont {Shen}, \citenamefont {Tao}, \citenamefont {Kim}, \citenamefont {Kn{\"u}ppel}, \citenamefont {Han}, \citenamefont {Zhang}, \citenamefont {Watanabe}, \citenamefont {Taniguchi}, \citenamefont {Chowdhury}, \citenamefont {Shan},\ and\ \citenamefont {Mak}}]{zhao24}%
  \BibitemOpen
  \bibfield  {author} {\bibinfo {author} {\bibfnamefont {W.}~\bibnamefont {Zhao}}, \bibinfo {author} {\bibfnamefont {B.}~\bibnamefont {Shen}}, \bibinfo {author} {\bibfnamefont {Z.}~\bibnamefont {Tao}}, \bibinfo {author} {\bibfnamefont {S.}~\bibnamefont {Kim}}, \bibinfo {author} {\bibfnamefont {P.}~\bibnamefont {Kn{\"u}ppel}}, \bibinfo {author} {\bibfnamefont {Z.}~\bibnamefont {Han}}, \bibinfo {author} {\bibfnamefont {Y.}~\bibnamefont {Zhang}}, \bibinfo {author} {\bibfnamefont {K.}~\bibnamefont {Watanabe}}, \bibinfo {author} {\bibfnamefont {T.}~\bibnamefont {Taniguchi}}, \bibinfo {author} {\bibfnamefont {D.}~\bibnamefont {Chowdhury}}, \bibinfo {author} {\bibfnamefont {J.}~\bibnamefont {Shan}},\ and\ \bibinfo {author} {\bibfnamefont {K.~F.}\ \bibnamefont {Mak}},\ }\bibfield  {journal} {\bibinfo  {journal} {Nature Physics}\ }\href {https://doi.org/10.1038/s41567-024-02636-4} {10.1038/s41567-024-02636-4} (\bibinfo {year} {2024})\BibitemShut {NoStop}%
\bibitem [{\citenamefont {Ciorciaro}\ \emph {et~al.}(2023)\citenamefont {Ciorciaro}, \citenamefont {Smole{\'n}ski}, \citenamefont {Morera}, \citenamefont {Kiper}, \citenamefont {Hiestand}, \citenamefont {Kroner}, \citenamefont {Zhang}, \citenamefont {Watanabe}, \citenamefont {Taniguchi}, \citenamefont {Demler},\ and\ \citenamefont {{\.I}mamo{\u g}lu}}]{ciorciaro23}%
  \BibitemOpen
  \bibfield  {author} {\bibinfo {author} {\bibfnamefont {L.}~\bibnamefont {Ciorciaro}}, \bibinfo {author} {\bibfnamefont {T.}~\bibnamefont {Smole{\'n}ski}}, \bibinfo {author} {\bibfnamefont {I.}~\bibnamefont {Morera}}, \bibinfo {author} {\bibfnamefont {N.}~\bibnamefont {Kiper}}, \bibinfo {author} {\bibfnamefont {S.}~\bibnamefont {Hiestand}}, \bibinfo {author} {\bibfnamefont {M.}~\bibnamefont {Kroner}}, \bibinfo {author} {\bibfnamefont {Y.}~\bibnamefont {Zhang}}, \bibinfo {author} {\bibfnamefont {K.}~\bibnamefont {Watanabe}}, \bibinfo {author} {\bibfnamefont {T.}~\bibnamefont {Taniguchi}}, \bibinfo {author} {\bibfnamefont {E.}~\bibnamefont {Demler}},\ and\ \bibinfo {author} {\bibfnamefont {A.}~\bibnamefont {{\.I}mamo{\u g}lu}},\ }\href {https://doi.org/10.1038/s41586-023-06633-0} {\bibfield  {journal} {\bibinfo  {journal} {Nature}\ }\textbf {\bibinfo {volume} {623}},\ \bibinfo {pages} {509} (\bibinfo {year} {2023})}\BibitemShut {NoStop}%
\bibitem [{\citenamefont {Haerter}\ and\ \citenamefont {Shastry}(2005)}]{haerter05}%
  \BibitemOpen
  \bibfield  {author} {\bibinfo {author} {\bibfnamefont {J.~O.}\ \bibnamefont {Haerter}}\ and\ \bibinfo {author} {\bibfnamefont {B.~S.}\ \bibnamefont {Shastry}},\ }\href {https://doi.org/10.1103/PhysRevLett.95.087202} {\bibfield  {journal} {\bibinfo  {journal} {Phys. Rev. Lett.}\ }\textbf {\bibinfo {volume} {95}},\ \bibinfo {pages} {087202} (\bibinfo {year} {2005})}\BibitemShut {NoStop}%
\bibitem [{\citenamefont {Sposetti}\ \emph {et~al.}(2014)\citenamefont {Sposetti}, \citenamefont {Bravo}, \citenamefont {Trumper}, \citenamefont {Gazza},\ and\ \citenamefont {Manuel}}]{sposetti14}%
  \BibitemOpen
  \bibfield  {author} {\bibinfo {author} {\bibfnamefont {C.~N.}\ \bibnamefont {Sposetti}}, \bibinfo {author} {\bibfnamefont {B.}~\bibnamefont {Bravo}}, \bibinfo {author} {\bibfnamefont {A.~E.}\ \bibnamefont {Trumper}}, \bibinfo {author} {\bibfnamefont {C.~J.}\ \bibnamefont {Gazza}},\ and\ \bibinfo {author} {\bibfnamefont {L.~O.}\ \bibnamefont {Manuel}},\ }\href {https://doi.org/10.1103/PhysRevLett.112.187204} {\bibfield  {journal} {\bibinfo  {journal} {Phys. Rev. Lett.}\ }\textbf {\bibinfo {volume} {112}},\ \bibinfo {pages} {187204} (\bibinfo {year} {2014})}\BibitemShut {NoStop}%
\bibitem [{\citenamefont {Kuhlenkamp}\ \emph {et~al.}(2024)\citenamefont {Kuhlenkamp}, \citenamefont {Kadow}, \citenamefont {Imamo\ifmmode~\breve{g}\else \u{g}\fi{}lu},\ and\ \citenamefont {Knap}}]{Kuhlenkamp24}%
  \BibitemOpen
  \bibfield  {author} {\bibinfo {author} {\bibfnamefont {C.}~\bibnamefont {Kuhlenkamp}}, \bibinfo {author} {\bibfnamefont {W.}~\bibnamefont {Kadow}}, \bibinfo {author} {\bibfnamefont {A.~m.~c.}\ \bibnamefont {Imamo\ifmmode~\breve{g}\else \u{g}\fi{}lu}},\ and\ \bibinfo {author} {\bibfnamefont {M.}~\bibnamefont {Knap}},\ }\href {https://doi.org/10.1103/PhysRevX.14.021013} {\bibfield  {journal} {\bibinfo  {journal} {Phys. Rev. X}\ }\textbf {\bibinfo {volume} {14}},\ \bibinfo {pages} {021013} (\bibinfo {year} {2024})}\BibitemShut {NoStop}%
\bibitem [{\citenamefont {Divic}\ \emph {et~al.}(2024)\citenamefont {Divic}, \citenamefont {Soejima}, \citenamefont {Crépel}, \citenamefont {Zaletel},\ and\ \citenamefont {Millis}}]{divic24}%
  \BibitemOpen
  \bibfield  {author} {\bibinfo {author} {\bibfnamefont {S.}~\bibnamefont {Divic}}, \bibinfo {author} {\bibfnamefont {T.}~\bibnamefont {Soejima}}, \bibinfo {author} {\bibfnamefont {V.}~\bibnamefont {Crépel}}, \bibinfo {author} {\bibfnamefont {M.~P.}\ \bibnamefont {Zaletel}},\ and\ \bibinfo {author} {\bibfnamefont {A.}~\bibnamefont {Millis}},\ }\href {https://arxiv.org/abs/2406.15348} {\bibinfo {title} {Chiral spin liquid and quantum phase transition in the triangular lattice hofstadter-hubbard model}} (\bibinfo {year} {2024}),\ \Eprint {https://arxiv.org/abs/2406.15348} {arXiv:2406.15348 [cond-mat.str-el]} \BibitemShut {NoStop}%
\bibitem [{\citenamefont {Anderson}\ \emph {et~al.}(2023)\citenamefont {Anderson}, \citenamefont {Fan}, \citenamefont {Cai}, \citenamefont {Holtzmann}, \citenamefont {Taniguchi}, \citenamefont {Watanabe}, \citenamefont {Xiao}, \citenamefont {Yao},\ and\ \citenamefont {Xu}}]{anderson23}%
  \BibitemOpen
  \bibfield  {author} {\bibinfo {author} {\bibfnamefont {E.}~\bibnamefont {Anderson}}, \bibinfo {author} {\bibfnamefont {F.-R.}\ \bibnamefont {Fan}}, \bibinfo {author} {\bibfnamefont {J.}~\bibnamefont {Cai}}, \bibinfo {author} {\bibfnamefont {W.}~\bibnamefont {Holtzmann}}, \bibinfo {author} {\bibfnamefont {T.}~\bibnamefont {Taniguchi}}, \bibinfo {author} {\bibfnamefont {K.}~\bibnamefont {Watanabe}}, \bibinfo {author} {\bibfnamefont {D.}~\bibnamefont {Xiao}}, \bibinfo {author} {\bibfnamefont {W.}~\bibnamefont {Yao}},\ and\ \bibinfo {author} {\bibfnamefont {X.}~\bibnamefont {Xu}},\ }\href {https://doi.org/10.1126/science.adg4268} {\bibfield  {journal} {\bibinfo  {journal} {Science}\ }\textbf {\bibinfo {volume} {381}},\ \bibinfo {pages} {325} (\bibinfo {year} {2023})}\BibitemShut {NoStop}%
\bibitem [{\citenamefont {K\"onig}\ \emph {et~al.}(2020)\citenamefont {K\"onig}, \citenamefont {Randeria},\ and\ \citenamefont {J\"ack}}]{koenig20}%
  \BibitemOpen
  \bibfield  {author} {\bibinfo {author} {\bibfnamefont {E.~J.}\ \bibnamefont {K\"onig}}, \bibinfo {author} {\bibfnamefont {M.~T.}\ \bibnamefont {Randeria}},\ and\ \bibinfo {author} {\bibfnamefont {B.}~\bibnamefont {J\"ack}},\ }\href {https://doi.org/10.1103/PhysRevLett.125.267206} {\bibfield  {journal} {\bibinfo  {journal} {Phys. Rev. Lett.}\ }\textbf {\bibinfo {volume} {125}},\ \bibinfo {pages} {267206} (\bibinfo {year} {2020})}\BibitemShut {NoStop}%
\bibitem [{\citenamefont {Peri}\ \emph {et~al.}(2024)\citenamefont {Peri}, \citenamefont {Ilani}, \citenamefont {Lee},\ and\ \citenamefont {Refael}}]{peri24}%
  \BibitemOpen
  \bibfield  {author} {\bibinfo {author} {\bibfnamefont {V.}~\bibnamefont {Peri}}, \bibinfo {author} {\bibfnamefont {S.}~\bibnamefont {Ilani}}, \bibinfo {author} {\bibfnamefont {P.~A.}\ \bibnamefont {Lee}},\ and\ \bibinfo {author} {\bibfnamefont {G.}~\bibnamefont {Refael}},\ }\href {https://doi.org/10.1103/PhysRevB.109.035127} {\bibfield  {journal} {\bibinfo  {journal} {Phys. Rev. B}\ }\textbf {\bibinfo {volume} {109}},\ \bibinfo {pages} {035127} (\bibinfo {year} {2024})}\BibitemShut {NoStop}%
\bibitem [{\citenamefont {Pichler}\ \emph {et~al.}(2024)\citenamefont {Pichler}, \citenamefont {Kadow}, \citenamefont {Kuhlenkamp},\ and\ \citenamefont {Knap}}]{pichler24}%
  \BibitemOpen
  \bibfield  {author} {\bibinfo {author} {\bibfnamefont {F.}~\bibnamefont {Pichler}}, \bibinfo {author} {\bibfnamefont {W.}~\bibnamefont {Kadow}}, \bibinfo {author} {\bibfnamefont {C.}~\bibnamefont {Kuhlenkamp}},\ and\ \bibinfo {author} {\bibfnamefont {M.}~\bibnamefont {Knap}},\ }\href {https://doi.org/10.1103/PhysRevB.110.045116} {\bibfield  {journal} {\bibinfo  {journal} {Phys. Rev. B}\ }\textbf {\bibinfo {volume} {110}},\ \bibinfo {pages} {045116} (\bibinfo {year} {2024})}\BibitemShut {NoStop}%
\bibitem [{\citenamefont {Takahashi}(1977)}]{takahashi77}%
  \BibitemOpen
  \bibfield  {author} {\bibinfo {author} {\bibfnamefont {M.}~\bibnamefont {Takahashi}},\ }\href {https://doi.org/10.1088/0022-3719/10/8/031} {\bibfield  {journal} {\bibinfo  {journal} {Journal of Physics C: Solid State Physics}\ }\textbf {\bibinfo {volume} {10}},\ \bibinfo {pages} {1289} (\bibinfo {year} {1977})}\BibitemShut {NoStop}%
\bibitem [{\citenamefont {Giamarchi}(2003)}]{giamarchi}%
  \BibitemOpen
  \bibfield  {author} {\bibinfo {author} {\bibfnamefont {T.}~\bibnamefont {Giamarchi}},\ }\href {https://doi.org/10.1093/acprof:oso/9780198525004.001.0001} {\emph {\bibinfo {title} {Quantum Physics in One Dimension}}}\ (\bibinfo  {publisher} {Oxford University Press},\ \bibinfo {year} {2003})\BibitemShut {NoStop}%
\end{thebibliography}%
\bibliographystyle{apsrev4-2}

\clearpage

\appendix

\section{perturbation calculation of Heisenberg exchange at $n=4/3$}
\label{app:perturb}
In the case of $\bar{n}=4/3$ filling at $V’=0$, our DMRG calculations indicate for $U \gg t$ that charges become localized, with singly-occupied sites forming an effective honeycomb lattice. Thus, we may fix a classical charge configuration and then derive exchange interactions between the spin-1/2 degrees of freedom mediated by virtual charge fluctuations by standard $t/U$ perturbation theory.
The unperturbed Hamiltonian $\mathcal{H}_0$ is taken to be the repulsive interaction term in Eq.~\eqref{eq:h-model}, and the perturbation is given by the hopping term $\mathcal{H}_t$, i.e.
\begin{subequations}
    \begin{align}
        \mathcal{H}_0&=\mathcal{H}_U+\mathcal{H}_V
        \\
        \mathcal{H}^{\prime}&=\mathcal{H}_t
        \equiv -t\sum_{ij \in \text{n.n.}}x_{ij}
    \end{align}
\end{subequations}
where the (directed) $x_{ij}$ are defined as
\begin{equation}
    x_{ij}=\sum_{\sigma}c_{i\sigma}^{\dagger}c_{j\sigma}^{\vphantom{\dagger}}.
\end{equation}
The perturbation is thus a sum over all directed bonds $x_{ij}$. 
We formulate the perturbative expansion in $t \ll U \sim V$ following the method of Takahashi \cite{takahashi77}, where the effective Hamiltonian up to fourth order in $t/U$ projected into the ground-state manifold (classical charge configuration) is given by
\begin{align}
    h_2
    &=P_0 \mathcal{H}_t \frac{\mathds{1}-P_0}{E_0-\mathcal{H}_0} \mathcal{H}_t P_0
    \label{eq:h2}
    \\
    h_3
    &=P_0 \mathcal{H}_t \frac{\mathds{1}-P_0}{E_0-\mathcal{H}_0} \mathcal{H}_t
    \label{eq:h3}
    \frac{\mathds{1}-P_0}{E_0-\mathcal{H}_0} \mathcal{H}_t P_0
    \\
    h_4
    &=
    P_0\mathcal{H}_t\frac{\mathds{1}-P_0}{E_0-\mathcal{H}_0}\mathcal{H}_t
    \frac{\mathds{1}-P_0}{E_0-\mathcal{H}_0}\mathcal{H}_t
    \frac{\mathds{1}-P_0}{E_0-\mathcal{H}_0}\mathcal{H}_tP_0
    \notag 
    \\
    &   -\frac{1}{2}\biggl(
    P_0\mathcal{H}_t\frac{\mathds{1}-P_0}{(E_0-\mathcal{H}_0)^2}\mathcal{H}_t
    P_0\mathcal{H}_t\frac{\mathds{1}-P_0}{E_0-\mathcal{H}_0}\mathcal{H}_tP_0
    \notag \\
    &+P_0\mathcal{H}_t\frac{\mathds{1}-P_0}{E_0-\mathcal{H}_0}\mathcal{H}_t
    P_0\mathcal{H}_t\frac{\mathds{1}-P_0}{(E_0-\mathcal{H}_0)^2}\mathcal{H}_tP_0
    \biggr) \label{eq:discontinuous}
\end{align}
Here, $E_0$ is the ground state energy of $\mathcal{H}_0$ and $P_0$ is the projection operator to the ground subspace of $\mathcal{H}_0$.
Each $\mathcal{H}_t$ contributes one directed bond $x_{ij}$.
The combination of $x_{ij}$s can be rewritten in spin operators. The relevant expressions read
\begin{widetext}
    \begin{subequations}\begin{align}
    &P_0x_{ij}x_{ji}P_0 = -2 \vec{S_i}\cdot \vec{S_j}+1/2 \quad
    \text{if $n_i=n_j=1$,}  \\
    &P_0x_{ik}x_{kj}x_{ji}P_0
    = 2\vec{S_j}\cdot\vec{S_k}+1/2 \quad \text{if $n_i=2,n_j=n_k=1$,} \\
    &P_0(x_{ik}x_{kj}x_{ji}+x_{ki}x_{ij}x_{jk})P_0
    = -2\vec{S_i}\cdot\vec{S_k}+1/2 \quad \text{if $n_i=n_j=n_k=1$,} \\
    &P_0(x_{Ok}x_{kj}x_{ji}x_{iO}+x_{Oi}x_{ij}x_{jk}x_{kO})P_0 = 2\vec{S_i}\cdot\vec{S_j}+2\vec{S_i}\cdot\vec{S_k}
    +2\vec{S_j}\cdot\vec{S_k}+1/2 \quad \text{if $n_O=2,n_i=n_j=n_k=1$,} \\
    &P_0(x_{ij}x_{ji}x_{jk}x_{kj}+x_{jk}x_{kj}x_{ij}x_{ji})P_0 = 2\vec{S_i}\cdot \vec{S_k}-2\vec{S_i}\cdot \vec{S_j}-2\vec{S_j}\cdot \vec{S_k}+1/2 \quad \text{if $n_i=n_j=n_k=1$.}
    \end{align}\end{subequations}
\end{widetext}
Combinations of $x_{ij}$ for other fillings give trivial constants. Here $n_i$, refers to the number of charges at site $i$ in the classical ground state charge configuration.
Counting all the possible combinations of $x_{ij}$ which survive the projections $P_0$ or $\mathds{1}-P_0$, and expressing the electron operators in spin operators, we arrive at the effective Hamiltonian
\begin{align}
    \mathcal H_{\text{eff}}
    &=J_1\sum_{\langle ij\rangle}\vec{S}_i\cdot \vec{S}_j
    +J_2\sum_{\langle \langle ij \rangle \rangle} \vec{S}_i\cdot \vec{S}_j
    +J_3\sum_{\langle \langle \langle ij \rangle \rangle \rangle}\vec{S}_i\cdot \vec{S}_j \nonumber\\
    &+K\sum{}'_{\langle ij \rangle,\langle kl\rangle}\left(\vec{S}_i\cdot \vec{S}_j\right)\left(\vec{S}_k\cdot \vec{S}_l\right) 
    .\label{eq:app_eff_H}
\end{align}
The coupling constants are given by
\begin{subequations}\begin{align}
    J_1/t^4
    &=
    -\frac{104}{(U-V)^3}
    +\frac{1}{(U-V)^2}
    \bigg(
    \frac{32}{U+V}+\frac{20}{U+2V}
    \notag \\
    &+\frac{44}{U}
    -\frac{112}{3V}+\frac{64}{2U-V}
    +\frac{64}{2U-3V}-\frac{16}{t}
    \bigg)
    \notag \\
    &+\frac{1}{U-V}
    \bigg(
    \frac{32}{(U+V)V}+\frac{20}{(U+2V)V}+\frac{12}{UV}
    -\frac{50}{3V^2}+ \frac{4}{t^{2}}
    \bigg)
    \notag \\
    &+\frac{1}{V^2}
    \bigg(
    \frac{8}{U+V}+\frac{5}{U+2V}+\frac{3}{U}
    +\frac{4}{U+3V}-\frac{7}{3V}-\frac{2}{t}
    \bigg)
    \label{eq:J1}
    \\
    J_2/t^4
    &=
    \frac{8}{(U-V)^3}
    -\frac{1}{(U-V)^2}
    \bigg(
    \frac{2}{U}
    +\frac{1}{V}
    \bigg)
    \notag \\
    &+\frac{1}{U-V}
    \bigg(
    \frac{2}{UV}-\frac{1}{V^2}
    \bigg)
    +\frac{1}{V^2}
    \bigg(
    \frac{2}{U+2V}+\frac{3}{2U}-\frac{1}{4V}
    \bigg)
    \label{eq:J2}
    \\
    J_3/t^4
    &=
    \frac{1}{V^2}
    \bigg(
    \frac{2}{U+2V}+\frac{1}{U}
    \bigg)
    \label{eq:J3}
    \\
    K/t^4
    &=
    \frac{32}{(U-V)^3}
    +\frac{32}{(U-V)^2}
    \bigg(
    \frac{1}{-2U+V}+\frac{1}{-2U+3V}
    \bigg).
    \label{eq:app_K}
\end{align}\end{subequations}


\section{$\Uone$ and $\SUtwo$ DRMG comparison}
\label{sec:U1vsSU2}
\begin{figure}[ht]
  \centering
  \includegraphics[width=0.9\linewidth]{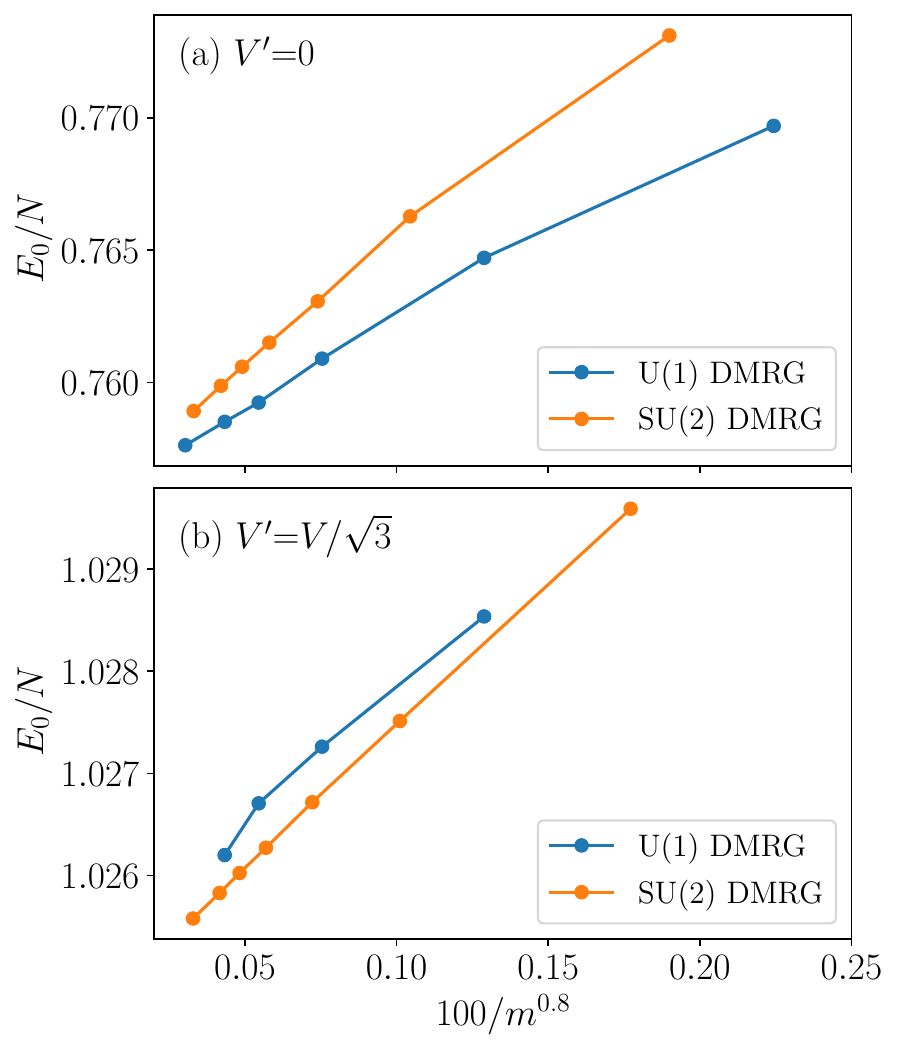}
  \caption{Comparison of ground state energies between $\SUtwo$ and $\Uone$ DMRG simulations at filling $\bar{n}=5/4$, for (a) $V'=0$ and (b) $V'=V/\sqrt{3}$.
  }
  \label{fig:U1vsSU2}
\end{figure}

In this Appendix, we compare the the ground-state energy for filling $\bar{n}=5/4$ in the cases of $V'=0$ or $V/\sqrt{3}$ as obtained from $\Uone$ and $\SUtwo$ preserving DMRG simulations.
As stated in Sec.~\ref{sec:dmrg}, the $\Uone$ DMRG enforces only a $\Uone$ spin rotation symmetry, and thus allows for a ground state with a spontaneously broken spin rotation symmetry. In contrast, the $\SUtwo$ DMRG enforces a full $\SUtwo$ spin rotation symmetry, as present in the model Hamiltonian, and thus converges to states without any spin polarization.

Fig.~\ref{fig:U1vsSU2} shows the energy per site calculated in $\Uone$ and $\SUtwo$ DMRG, where the bond dimension of the $\SUtwo$ multiplets is converted to an equivalent $\Uone$ bond dimension.
We see that, for $V'=0$, the $\Uone$ DMRG always gives a lower ground state, while the $\SUtwo$ DMRG gives a lower energy for $V'=V/\sqrt{3}$.
This is in accord with our results in Sec.~\ref{sec:R54_Vp0} and Sec.~\ref{sec:kagome-crystal}, where a spin-density wave is observed for $V'=0$, and a quantum paramagnet on an emergent Kagom\'e charge order is found for $V'=V/\sqrt{3}$.

\section{Bosonization analysis for $\bar{n}=5/4$ with $V'=0$}
\label{app:bosonization}
In this Appendix we seek to understand the gapped behavior at filling $\bar{n}=5/4$ with $V'=0$, as discussed in Sec.~\ref{sec:R54_Vp0}. 
On the finite-width cylinder geometry, when the length is long compared to the cylinder circumference, we may regard the system to be quasi--one-dimensional.
Then, we use standard bosonization arguments to analyse how interactions can lead to additional symmetry breaking. 

As a first step, we treat the $\sqrt{3}\times \sqrt{3}$ CDW as a background potential, and assume the $\SUtwo$ spin rotation symmetry is unbroken, consistent with the de facto one-dimensionality of the system.
A corresponding mean-field Hamiltonian then reads
\begin{equation} \label{eq:mft-cdw}
    \mathcal H=\mathcal H_t+\mathcal H_\rho
    =-t\sum_{ij,\sigma}(c_{i,\sigma}^{\dagger}c_{j,\sigma}^{\vphantom{\dagger}}+\hc)
    +\sum_{i,\sigma} \rho_i c_{i,\sigma}^{\dagger}c_{i,\sigma}^{\vphantom{\dagger}}
\end{equation}
where the potential term $\mathcal H_{\rho}$ accounts for the CDW order imposed by Coulomb repulsion, and we absorb any coupling constants into the amplitude of $\rho_i$. 
With the unit vectors for the triangular lattice given by $\hat{x} = (1,0)$ and $\hat{y} = (1/2,\sqrt{3}/2)$, we consider periodic boundary conditions along both $\hat{x}$ (length) and $\hat{y}$ (width) directions, with a circumference of $L_y=6$, and we take the long cylinder limit ($L_x\to \infty$).
The mean field $\rho_i$ models a $\sqrt{3}\times \sqrt{3}$ CDW pattern as
\begin{equation}
    \rho_i=\rho_0 \text{Re}[\eu^{\iu \bvec{K} \cdot \mathbf r_i}]
    \quad \text{with}
    \quad \mathbf K=\left(\frac{4\pi}{3},0\right)
    \label{app_eq:CDW}
\end{equation}
Note that for convenience we choose a different $\bvec{K}$-point compared to the main text, but they are related by a $C_3$ rotation and therefore lead to the same charge modulation.
Then, we Fourier transform the fermionic operators in \eqref{eq:mft-cdw}, labelling the wavenumbers (i.e. projections of the 2D wavevector) along $\hat{x}$ and $\hat{y}$ by $k$ and $q$, respectively. 
Since $L_y=6$, $q$ can only take 6 discrete values mod $2\pi$: $\{0, \frac{2\pi}{6},\frac{4\pi}{6},\frac{6\pi}{6},\frac{8\pi}{6},\frac{10\pi}{6}\}$, while in the long-cylinder limit $k\in [-\pi,\pi)$.

When the CDW pattern as given in Eq.~\eqref{app_eq:CDW} is turned on, a state with wavenumbers $(k,q)$ can be scattered to $(k+\frac{2\pi}{3},q-\frac{2\pi}{3})$ or $(k-\frac{2\pi}{3},q+\frac{2\pi}{3})$.
Therefore, the Brillouin zone should be folded along both directions, and the truly independent $(k,q)$ takes values in $k\in [-\frac{\pi}{3},\frac{\pi}{3})$ and $q\in \{0,\frac{2\pi}{6}\}$.
In momentum space, $\mathcal H_\rho$ can be written as 
\begin{align}
    \mathcal H_\rho=\rho_0 \sum_{k,q}{}^{'}
    \bigl(c^{\dagger}_{k+\frac{2\pi}{3},q-\frac{2\pi}{3}}
    c_{k,q}^{\vphantom{\dagger}}
    +c^{\dagger}_{k-\frac{2\pi}{3},q+\frac{2\pi}{3}}
    c_{k,q}^{\vphantom{\dagger}}
    \bigr)
\end{align}
where $\sum^{'}_{k,q}$ means the summation is over the restriced $k$ and $q$ values, and $k,q$ in the subscripts are mod $2\pi$. 
Here, we suppressed the spin indices for notational simplicity.
After the Brillouin zone folding, the original Brillouin zones becomes partitioned into distinct sectors.
The corresponding spinor with wavenumbers $(k,q)$ in the restricted Brillouin zone as
\begin{equation}
    \Psi_{k,q}
    =\begin{pmatrix}
        c_{k,q} & c_{k,q+\frac{2\pi}{3}}
        & c_{k,q+\frac{4\pi}{3}} & c_{k+\frac{2\pi}{3},q}
        & \dots
        & c_{k+\frac{4\pi}{3},q+\frac{4\pi}{3}}
    \end{pmatrix}
    \label{app_eq:spinor}
\end{equation}
For each $(k,q)$, the Hamiltonian is a 9 by 9 matrix. $\mathcal H_t$ is still diagonal, and $\mathcal H_\rho$ reads
\begin{align}
    \mathcal H_\rho=
    \begin{pmatrix}
        0 & 0 & 0 & 0 & 0 & \rho & 0 & \rho & 0 \\
        0 & 0 & 0 & \rho & 0 & 0 & 0 & 0 & \rho \\
        0 & 0 & 0 & 0 & \rho & 0 & \rho & 0 & 0 \\
        0 & \rho & 0 & 0 & 0 & 0 & 0 & 0 & \rho \\
        0 & 0 & \rho & 0 & 0 & 0 & \rho & 0 & 0 \\
        \rho & 0 & 0 & 0 & 0 & 0 & 0 & \rho & 0 \\
        0 & 0 & \rho & 0 & \rho & 0 & 0 & 0 & 0 \\
        \rho & 0 & 0 & 0 & 0 & \rho & 0 & 0 & 0 \\
        0 & \rho & 0 & \rho & 0 & 0 & 0 & 0 & 0 \\
    \end{pmatrix}
\end{align}
In the reduced Brillouin zone, $q\in \{0, \frac{2\pi}{6}\}$ can take two possible values. In the spirit of the effective one-dimensional geometry, we can plot a one-dimensional band structure as a function of the wavenumber $k \in [-\pi/3,\pi/3)$ along the cylinder.
In total, there will be  $2 \times 9=18$ bands for each $k$.

\begin{figure}[ht]
  \centering
  \includegraphics[width=0.9\linewidth]{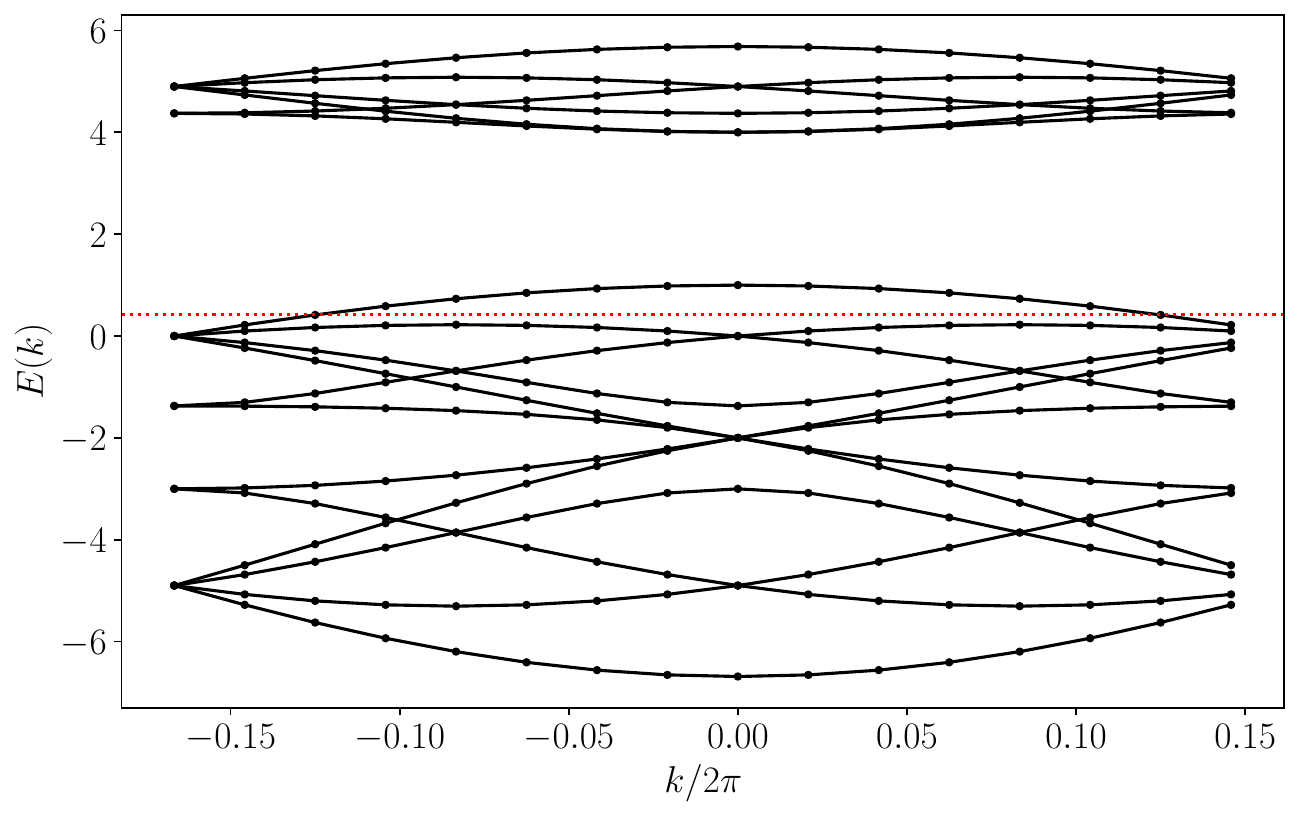}
  \caption{One-dimensional band structure for a finite mean-field CDW background with amplitude $\rho_0 = 2 t$, as a function of wavenumber $k$ along the cylinder direction, with the transverse (rung) wavenumber corresponding to distinct bands. The red dotted line indicate the Fermi energy for $\bar{n}=5/4$ filling, which crosses a single band. Here, we have chosen a cylinder length of $L_x = 48$ sites and circumference $L_y = 6$, in accordance with the geometry studied in our DMRG simulations.
  }
  \label{fig:R54Vp0bands}
\end{figure}
We numerically diagonalize the quadratic Hamiltonian in the basis of Eq.~\eqref{app_eq:spinor}, for each $k$ and $q=0,\frac{2\pi}{6}$, and plot the bands along $k$ in the restricted Brillouin zone in Fig.~\ref{fig:R54Vp0bands}. 
When the CDW pattern is turned on with a moderate $\rho_0$, the Fermi energy (red dotted line) corresponding to filling $\bar{n}=5/4$ crosses only a single band.
Here we choose $\rho_0/t=2$, approximately matching the DRMG calculation as shown in Fig.~\ref{fig:R54Vp0} in the main text (after taking both $U$ and $V$ into consideration). The fact that the Fermi level crosses a single band remains robust under varying $\rho_0/t$.
Counting the filled states, we further deduce the corresponding Fermi wavevector $k_\mathrm{F} = \pi/4$.

In the convention of the spinor basis Eq.~\eqref{app_eq:spinor}, we can numerically get the weights of states $\psi(k)$ near the Fermi level corresponding to filling $\bar{n} = 5/4$ in terms of the original $c_{k,q}$ basis 
\begin{equation}
    \psi(k) = \alpha_1(k) c_{k,0} +\alpha_6(k) c_{k+\frac{2\pi}{3},-\frac{2\pi}{3}}+\alpha_8(k)c_{k-\frac{2\pi}{3},\frac{2\pi}{3}}. \label{app_eq:fermi_state}
\end{equation}
This is a superposition of the states that can be scattered into each other by $H_\rho$, and only involves components with $q=0, \pm \frac{2\pi}{3}$ components.

Next, we work out how the lattice symmetry acts on those states near the Fermi level.
With the $\sqrt{3}\times \sqrt{3}$ CDW background, the Hamiltonian is invariant under translations which act on the fermionic operators on site $i$ with coordinates $\bvec{r}_i$ as $T_1^{(\mathrm{CDW})}: c(\bvec{r}_i)\rightarrow c(\bvec{r}_i+\hat{x}+\hat{y})$ and $T_2^{(\mathrm{CDW})}: c(\bvec{r}_i)\rightarrow c(\bvec{r}_i+2\hat{x}-\hat{y})$.
In momentum space, these translations act as
\begin{equation}
    T_1^{(\mathrm{CDW})}: c_{k,q}\rightarrow c_{k,q}\eu^{\iu(k+q)},
    \quad
    T_2^{(\mathrm{CDW})}: c_{k,q}\rightarrow c_{k,q} \eu^{\iu(2k-q)}
\end{equation}
According to Eq.~\eqref{app_eq:fermi_state}, the state at the Fermi level transforms under the lattice translation $T_1^{(\mathrm{CDW})}$ (with $\bvec{r}_i\rightarrow \bvec{r}_i+\hat{x}+\hat{y}$) as
\begin{widetext}
\begin{align}
    \psi(k_\mathrm{F})&\rightarrow 
    \alpha_1(k_\mathrm{F})\eu^{\iu k_\mathrm{F}}c_{k_\mathrm{F},0}+
    \alpha_6(k_\mathrm{F})\eu^{\iu(k_\mathrm{F}+\frac{2\pi}{3}-\frac{2\pi}{3})}c_{k_\mathrm{F}+\frac{2\pi}{3},-\frac{2\pi}{3}}+
    \alpha_8(k_\mathrm{F})\eu^{\iu (k_\mathrm{F}-\frac{2\pi}{3}+\frac{2\pi}{3})}c_{k_\mathrm{F}-\frac{2\pi}{3},\frac{2\pi}{3}}
    \notag \\
    &=\eu^{\iu k_\mathrm{F}}\psi(k_\mathrm{F})
    \label{app_eq:Rferm_trans}
\end{align}
\end{widetext}
And similarly,
\begin{equation}
    T_2^{(\mathrm{CDW})}:\psi(-k_\mathrm{F})\rightarrow \eu^{-\iu k_\mathrm{F}}\psi(-k_\mathrm{F})
    \label{app_eq:Lferm_trans}
\end{equation}
These transformations under translation operations constrain the possible interaction terms in the bosonized theory, as we will see soon.

At lowest energies, we may focus on the single band that crosses the Fermi level. We consider interactions among these degrees of freedom, which we treat with a standard bosonization procedure for spinful electrons \cite{giamarchi},
\begin{equation}
    \psi_{r,\sigma}(x)\sim U_{r,\sigma}\eu^{-\iu r k_\mathrm{F} x}
    \eu^{-\frac{\iu}{\sqrt{2}}[r\phi_{\rho}(x)-\theta_{\rho}(x)
    +\sigma(r\phi_{\sigma}(x)-\theta_{\sigma}(x))]}
\end{equation}
where $U_{r,\sigma}$ is the Klein factor that has no spatial dependence and endows the $\psi_{r,\sigma}$ with fermionic statistics.
Here, $x$ is a coordinate along the cylinder's long axis, $r$ denotes the left ($r=-1$) or right $(r=1)$ mover, and the minus sign in the first exponential due to the fact that the band dispersion has negative slope at $+k_\mathrm{F}$. $\sigma = \uparrow,\downarrow$ denotes spin, and $\phi_{\rho}$ and $\phi_{\sigma}$ are defined as
\begin{align}
    \phi_{\rho}&=\frac{\phi_{\uparrow}+\phi_{\downarrow}}{\sqrt{2}} \notag \\
    \phi_{\sigma}&=\frac{\phi_{\uparrow}-\phi_{\downarrow}}{\sqrt{2}}
\end{align}
describing the charge sector and spin sector, respectively, and $\phi_{\uparrow}$ and $\phi_{\downarrow}$ are standard bosonic operators for spin up and down species. $\theta_{\rho}$ and $\theta_{\sigma}$ are defined similarly.

Under the translation Eq.~\eqref{app_eq:Rferm_trans} and \eqref{app_eq:Lferm_trans}, the bosonic operators translate as 
\begin{align}
    \phi_{\rho}\rightarrow \phi_{\rho}+\frac{\sqrt{2}}{4}\pi
\end{align}
Further, the $\Uone$ charge conservation and $\Uone$ spin-rotation symmetry (about the $\hat{z}$-axis), acting as $\psi_{\sigma}\rightarrow \psi_{\sigma}\eu^{i\delta \theta_\rho}$ and $\psi_{\sigma}\rightarrow \psi_{\sigma}\eu^{i\sigma\delta \theta_\sigma}$, are implemented on the bosonic fields as
\begin{align}
    \theta_\rho & \rightarrow \theta_\rho+\sqrt2 \delta \theta_\rho
    \notag \\
    \theta_\sigma & \rightarrow \theta_\sigma+\sqrt{2}\delta\theta_\sigma    
\end{align}
Finally, the Hamiltonian possesses a parity symmetry $\psi \rightarrow -\psi$ (a subgroup of the $\Uone$ symmetry), which can be implemented via the bosonic fields as
\begin{align}
    \phi_{\rho(\sigma)}
    \rightarrow \phi_{\rho(\sigma)}
    +\frac{\sqrt{2}}{2}\pi,
\end{align}
or with a similar transformation for $\theta_{\rho(\sigma)}$.
The effective bosonic Hamiltonian should therefore be invariant under the above transformations.

The usual quadratic bosonic Hamiltonian reads
\begin{equation}
    \mathcal H^0=\mathcal H_\rho^0+\mathcal H_\sigma^0
\end{equation}
with 
\begin{align}
    \mathcal H_{\rho}^0&=\frac{1}{2\pi}\int dx\, u_{\rho}K_{\rho}(\pi\Pi_\rho(x))^2+\frac{u_{\rho}}{K_{\rho}}(\nabla\phi_\rho(x))^2
    \notag \\
    \mathcal H_{\sigma}^0&=\frac{1}{2\pi}\int dx\, u_{\sigma}K_{\sigma}(\pi\Pi_\sigma(x))^2+\frac{u_{\sigma}}{K_{\sigma}}(\nabla\phi_\sigma(x))^2    
\end{align}
We now consider symmetry-allowed interactions on top of the CDW mean-field ground state that can lead to symmetry breaking.
The terms allowed by the above symmetry constraints can be
\begin{equation}
    \mathcal H_{\text{int}}=g_{\sigma}\cos(2\sqrt{2}\phi_\sigma)+
    g_{\rho}\cos(4\sqrt2\phi_\rho) + \dots,
    \label{app_eq:int_term}
\end{equation}
where we omit higher harmonics of the fields.

These interactions can also be inferred from a fermionic treatment as follows:
The $\cos(2\sqrt{2}\phi_{\sigma})$ for the spin sector arises from the spin-flipping scattering process between left and right movers with opposite spins, i.e.$\psi_{L,\sigma}^{\dagger}\psi_{R,-\sigma}^{\dagger}\psi_{L,-\sigma}\psi_{R,\sigma}$. 
The Umklapp scattering can generate the charge sector sine-Gordon term $\cos(4\sqrt{2}\phi_{\rho})$: considering $k_\mathrm{F}=\pi/4$ as calculated above, scattering processes of the form $\psi_R^{\dagger}\psi_R^{\dagger}\psi_R^{\dagger}\psi_R^{\dagger}\psi_L\psi_L\psi_L\psi_L$ are permitted since the momentum transfer is $\pi/4\times 2\times4=2\pi$, and momentum is conserved up to $2\pi$ in the presence of a lattice.

If a cosine-like term as appearing in \eqref{app_eq:int_term} is relevant, then it would fix the field ($\phi_\sigma$ or $\phi_\rho$) to its respective minimizing value: excitations about this minimum are massive and therefore corresponding sector would be gapped.
The sign of the coefficient of the cosine-like term determines the ordered value and dictates the behaviors of the correlation functions. 

From the DMRG calculation, we infer that both charge and spin sectors are gapped, and therefore the sine-Gordon terms must be relevant.
To determine the signs of the coefficients without diving into a full renormalization group calculation, we proceed with a a classical limit, where we neglect any fluctuations of the fields $\phi_\sigma$ and $\phi_\rho$ and determine which minima of the interactions in \eqref{app_eq:int_term} are consistent with the order observed in the DMRG simulations. 
In terms of the bosonic operators, the charge density $\rho$, spin density $S^z$, and bond kinetic energy $b\equiv \sum_{\sigma}(\psi^{\dagger}_{\sigma}(x)\psi_{\sigma}(x+1)+\hc)$ can be expressed as
\begin{widetext}
\begin{align}
    \rho(x)
    &=\rho_0-\frac{\sqrt{2}}{\pi}\nabla\phi_\rho(x)
    +\rho_2\cos{(2k_\mathrm{F}x-\sqrt{2}\phi_\rho(x))}\cos(\sqrt{2}\phi_\sigma(x))
    \notag \\
    &+\rho_4\cos(4k_\mathrm{F}x-2\sqrt{2}\phi_\rho(x))
    +\rho_8\cos(8k_\mathrm{F}x-4\sqrt{2}\phi_\rho(x))
    +\dots
    \\
    S^z(x)
    &=-\frac{\sqrt{2}}{\pi}\nabla\phi_{\sigma}(x)
    +s_2\cos(2k_\mathrm{F} x-\sqrt{2}\phi_\rho(x))\sin(\sqrt{2}\phi_\sigma)
    \notag \\
    &+s_4\cos(4k_\mathrm{F} x-2\sqrt{2}\phi_\rho)\sin(2\sqrt{2}\phi_\sigma)
    +\dots
    \\
    b(x)
    &=b_0-\frac{2\sqrt{2}}{\pi}\nabla\phi_\rho(x)
    +b_2\cos(2k_\mathrm{F}x-\sqrt{2}\phi_\rho(x)+k_\mathrm{F})\cos(\sqrt{2}\phi_\sigma(x))
    \notag \\
    &+b_4\cos(4k_\mathrm{F}x-2\sqrt{2}\phi_\rho(x)+2k_\mathrm{F})
    +b_8\cos(8k_\mathrm{F}x-4\sqrt{2}\phi_\rho(x)+4k_\mathrm{F})
    +\dots
    \label{eq:rho_in_b}
\end{align}
\end{widetext}
where the $\cos(2\sqrt{2}\phi_{\sigma}(x))$ that seemingly comes along with the $\eu^{\iu (4k_\mathrm{F}x)}$ term can be eliminated with the $\cos(2\sqrt{2}\phi_{\sigma}(x))$ term in $\mathcal H_{\sigma}$ under a renormalization group step \cite{giamarchi}.
Note that the precise values above are abitrary and can (in principle) be determined via a full renormalization group calculation. 
\begin{figure}[ht]
  \centering
  \includegraphics[width=0.9\linewidth]{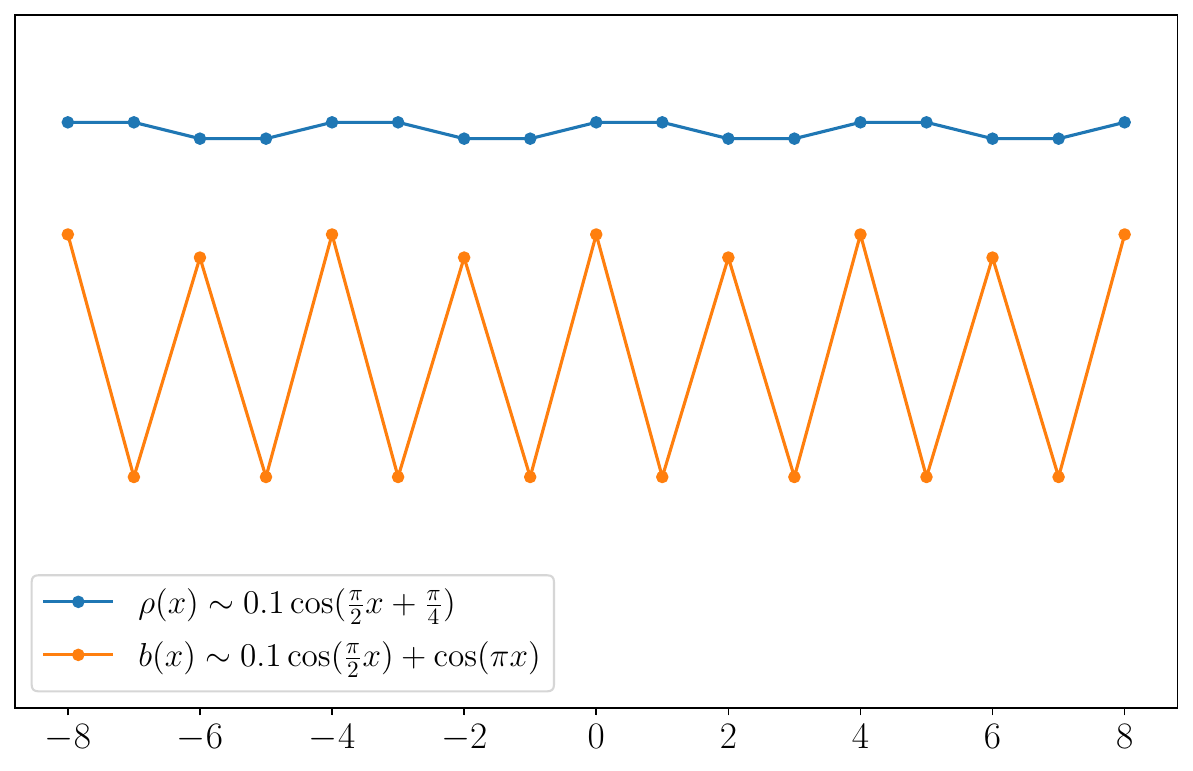}
  \caption{Expectation values of the density $\rho(x)$ and kinetic bond energy $b(x)$ in the classical limit of the bosonic Hamiltonian. The coefficients are chosen so that the patterns resemble the DMRG results. Careful renormalization group calculations are needed to get the precise values of coefficients.
  }
  \label{fig:boson_classical}
\end{figure}

For consistency with results of our DMRG calculations shown in Fig.~\ref{fig:R54Vp0_EE}, we now require that $S^z(x)$ vanishes (preserving $\SUtwo$ spin rotation symmetry) and a period-2 structure of the bond kinetic energy $b(x)$: this leads to a scenario where $g_\sigma<0$ and $g_\rho>0$, with coefficients $b_2\ll b_4$.
In this case, the energy Eq.~\eqref{app_eq:int_term} is minimized by the field configurations $\phi_\sigma=0$ and $\phi_\rho=\frac{\sqrt{2}}{8}\pi$.
Consequently, the charge density $\rho(x)\sim \rho_2\cos(\frac{\pi}{2}x+\frac{\pi}{4})$, $b(x)\sim b_2\cos(\frac{\pi}{2}x)+b_4\cos(\pi x)$, and $S^z(x)=0$ as expected.
Such a pattern of $\rho(x)$ and $b(x)$ are shown in Fig.~\ref{fig:boson_classical}, where we heuristically choose some coefficients so that the patterns resemble the DMRG results. 
Significantly, we stress that the contribution $\cos(\pi x)$ leads to a dimerization pattern, and thus a doubling of the unit cell, which at the considered filling $\bar{n}=5/4$ hosts an integer number of electrons and permits insulating behaviour.

We would like to comment that our bosonization analysis is, strictly speaking, appropriate for a weak coupling scenario. While the the case of $U/t=12$ of interest to us lies outside such a weak-coupling analysis, the symmetry-based analysis allows us to analyse possible scenarios of interaction-induced additional (discrete) symmetry breaking (on top of the CDW ordering).
The obtained results qualitatively match the numerical simulations.

\end{document}